\documentclass{ws-ijmpa}
\usepackage[super,compress]{cite}
\usepackage{graphicx}
\begin{document}
\markboth{M. M. Almarashi}{The second lightest CP-even Higgs boson signals in the NMSSM at the LHC}
\catchline{}{}{}{}{}
%
%%%%%%%%%%%%%%%%%%%%%%%%%%%%%%%%%%%%%%%%%%%%%%%%%%%%%%%%%%%%%%%%%%%%

\title{The second lightest CP-even Higgs boson signals in the NMSSM at the LHC}
\author{M. M. Almarashi}
\address{Physics Department, Faculty of Science, Taibah University\\
P. O. Box 344, Al-Madinah Al-Munawwarah, Kingdom of Saudia Arabia\\
al\_marashi@hotmail.com}

\maketitle

\begin{history}
\received{Day Month Year}
\revised{Day Month Year}
\end{history}

\begin{abstract}
We study the signal rates of the second lightest CP-even Higgs boson, $h_2$, of the NMSSM produced in gluon fusion,
in association with bottom quarks
and in association with top quarks, which is not the SM-like Higgs boson, at the LHC. We evaluate the production rates of the $h_2$ 
in the SM fermionic and bosonic final states in addition to
 $a_1a_1$, $h_1h_1$ and $Za_1$ final states. It is observed that the size of the signal rates 
in some regions of the NMSSM parameter space is quite large and that could help extracting the $h_2$ signals at the LHC 
through a variety of decay channels.
\end{abstract}

\ccode{PACS numbers: 14.80.Da, 12.60.Fr, 14.80.Ec}

\section{Introduction}
\label{sect:intro}

Assuming CP-conservation in the Higgs sector, the Next-to-Minimal Supersymmetric Standard Model
(NMSSM) \cite{NMSSM1,NMSSM2,NMSSM3,NMSSM4,NMSSM5,NMSSM6,NMSSM7,NMSSM8,NMSSM9,NMSSM10,NMSSM11} 
predicts one or more light CP-even Higgs boson, either the lightest Higgs boson $h_1$ or the second lightest Higgs boson $h_2$, which could
be the Higgs particle with a mass around 125 GeV discovered by the Large Hadron Collider (LHC) \cite{CMS1-Higgs,ATLAS1-Higgs,CMS2-Higgs,ATLAS2-Higgs}.
This discovered SM-like Higgs can be accommodated in the context of the NMSSM
without much fine tuning, see
Refs. ~\refcite{NMSSMrecent1}--\refcite{Das:2017tob}
for more details.

Within the NMSSM the Higgs fields consists of the two MSSM-type Higgs
doublets in addition to the Higgs singlet,
giving rise to seven Higgs states: three CP-even Higgses $h_{1, 2, 3}$ ($m_{h_1} < m_{h_2} < m_{h_3}$), 
two CP-odd Higgses $a_{1, 2}$ ($m_{a_1} < m_{a_2} $) and a pair of charged Higgses $h^{\pm}$. When 
the singlet field acquires a vacuum expectation value, an `effective' $\mu$-term, $\mu_{\rm eff}$, is
generated automatically, which is naturally of the order of the electroweak scale \cite{NMSSM4}.
Moreover, due to the presence of the singlet field, the phenomenology of the NMSSM Higgs sector is richer than that of the MSSM, which has only
five Higgs states: $h$, $H$, $A$ and $H^{\pm}$. The phenomenology of the NMSSM Higgs sector has been studied by many authors, for a review and details see Refs.
~\refcite{Ellwanger:2011sk}--\refcite{Baum:2017gbj} and references therein. For the latest studies about the NMSSM Higgs sector
see Refs. ~\refcite{Beskidt:2017dil}--\refcite{Almarashi:2018ign}.

Following the discovery of the SM-like Higgs boson at the LHC, looking for other Higgs bosons at the LHC, if they exist, 
would point to the existence of supersymmetric models. Searches for additional heavy neutral Higgs bosons
have been carried out by the ATLAS and CMS Collaborations in the $b\bar b$ \cite{Chatrchyan:2013qga,Khachatryan:2015tra},
$\tau\tau$ \cite{Chatrchyan:2011nx,Chatrchyan:2012vp,Aad:2012cfr,Khachatryan:2014wca,Aad:2014vgg,Aaboud:2016cre,Aaboud:2017sjh,Sirunyan:2018zut}
and gauge boson, $WW$ and $ZZ$, \cite{Khachatryan:2015cwa,Aad:2015kna} final states. No signal has been found so far,
putting some constraints on such Higgs bosons.

In this paper, we study the production rates of  
 the second lightest Higgs boson $h_2$, if it is not the SM-like Higgs boson,
 of the NMSSM at the LHC through the following production channels: gluon fusion 
$gg\to h_2$ and associated production with bottom-antibottom pair $gg\to b\bar bh_2$ and top-antitop pair $gg\to t\bar th_2$ for 
the most promising final states. We calculate the inclusive cross section of the $h_2$. It is observed that there exist some regions of the NMSSM parameter space where the cross sections times branching
ratios are quite sizable and could help extracting the $h_2$ signals at the LHC through the above production
channels.

The paper is organized as follows. In section ~\ref{sect:Higgs} we briefly discuss the NMSSM Higgs sector, describing the NMSSM parameter 
space scans performed and the allowed decay channels of the $h_2$. In section ~\ref{sect:rates} we present 
the inclusive event rates of $h_2$ production
at the LHC for various decay channels.
Finally, we summarize our results in section ~\ref{sect:summa}.

\section{\ The Higgs sector of the NMSSM}
\label{sect:Higgs}

The scale invariant superpotential of the NMSSM, including only the third generation fermions, is given by
 \begin{equation}  
 W=h_t\hat Q\hat H_u\hat t^c_R - h_b\hat Q\hat H_d\hat b^c_R -h_{\tau}\hat L\hat H_d\hat {\tau}^c_R+\lambda\hat S\hat H_u\hat H_d
+\frac{1}{3}\kappa{\hat S}^3,
\end{equation}
where $ h_t$, $ h_b$, $ h_{\tau}$, $\lambda$ and $\kappa$ are dimensionless couplings.
The term $\lambda\hat S\hat H_u\hat H_d$ is required 
to solve the $\mu$-problem \cite{Kim:1983dt}. The last term, which is cubic in the singlet superfield,
has been introduced 
to break the Peccei-Quinn (PQ) symmetry \cite{Peccei:1977hh, Peccei:1977ur}.

The soft SUSY-breaking potential containing only the Higgs fields is given by

\begin{equation}
V_{\rm soft}=m_{H_u}^2|H_u|^2+m_{H_d}^2|H_d|^2+m_{S}^2|S|^2
             +\left(\lambda A_\lambda S H_u H_d + \frac{1}{3}\kappa A_\kappa S^3 + {\rm h.c.}\right),
\end{equation}
where $A_\lambda$ and $A_\kappa$ are dimensionful parameters of the order of the mass scale  of supersymmetric particles.

The NMSSM Higgs sector is fully described by six independent parameters: $\kappa$, $A_\kappa$, $\lambda$, $A_\lambda$, 
tan$\beta$ (the ratio of the vacuum expectation values (VEVs) of the two Higgs doublets, $H_u$ and $H_d$)
and $\mu_{\rm eff} = \lambda\langle S\rangle$
(where $\langle S\rangle$ is the VEV of the singlet field).

For our study of the NMSSM Higgs sector
we have used the package NMSSMTools5.1.2 \cite{NMHDECAY1,NMHDECAY2,NMSSMTools} to compute the masses, couplings and decay widths of all the NMSSM
Higgs bosons. The package systematically takes into account various theoretical and experimental constraints. 
We use the package to scan over some regions of the NMSSM parameter space in order to obtain a general view of the  phenomenology of
the second lightest CP-even Higgs boson, $h_2$, at the LHC. We have set the six tree level parameters in the following ranges:

\begin{center}
$0.001 \leq \lambda \leq  0.1$, \phantom{aa} $-0.1 \leq \kappa \leq  0.1$,\phantom{aa} $1.6 \leq \tan\beta \leq  60$, \phantom{aa} \\
$100 \leq \mu \leq  1000$ GeV, \phantom{aa} $-1000 \leq A_{\lambda} \leq  1000$ GeV,\phantom{aa} $-1000 \leq A_{\kappa} \leq  1000$ GeV. \\
\end{center}
Notice that we focus here on scenario with small values of $\lambda$ and $\kappa$ in order to increase 
the probability of getting a lot of surviving data points with $m_{h_2} \lesssim 500$ GeV. In such scenario, the PQ symmetry
is slightly broken, \footnote{The PQ symmetry is slightly broken when $\kappa$ is small , i.e. $\kappa << 1$.} which is favored
by the renormalization group flow from
the grand unification scale down to the electroweak scale, see Ref. ~\refcite{MNZ} for more details . Further information about the effect of
$\lambda$ and $\kappa$ on the two lightest CP-even Higgs boson masses can be obtained from the following sum rule \cite{MNZ}:
\begin{eqnarray}
m_{h_1}^2 + m_{h_2}^2 \approx M_Z^2 + \frac{1}{2}\kappa \langle S\rangle(4\kappa \langle S\rangle+\sqrt{2}A_\kappa)
\end{eqnarray}
where $\langle S\rangle=\sqrt{2}\mu/ \lambda$.

Remaining terms, contributing at higher order level, which are soft SUSY breaking right- and left-handed masses for the first
two generations and the third generation, 
soft SUSY breaking trilinear couplings and gaugino soft SUSY breaking masses
have been set as:\\
$\bullet\phantom{a}m_Q = m_U = m_D = m_L = m_E = m_{Q_3} = m_{U_3} = m_{D_3} = m_{L_3} = m_{E_3} = 1000$ GeV,\\
$\bullet\phantom{a}A_{U_3} = A_{D_3} = A_{E_3} = 2000$ GeV,\\
$\bullet\phantom{a} M_1 = 250$ GeV, $M_2 = 500$ GeV,  $M_3 = 1500$ GeV.\\

In our study for looking for the $h_2$ at the LHC, we focus on the following decay  
channels:
\begin{eqnarray*}
&&h_2\rightarrow\tau^+\tau^-,\phantom{aaa}h_2\rightarrow
b\bar b,\phantom{aaa}h_2\rightarrow t\bar t, \\ 
&&h_2\rightarrow\gamma\gamma,\phantom{aaa}h_2\rightarrow
Z\gamma,\phantom{aaa}h_2\rightarrow W^+W^-,\phantom{aaa}h_2\rightarrow ZZ, \\
&&h_2\rightarrow a_1a_1,\phantom{aaa}h_2\rightarrow h_1h_1,\phantom{aaa}h_2\rightarrow Za_1.
\end{eqnarray*}

We  perform a random scan  over some fraction of the NMSSM parameter space. The output of the scan contains masses, branching ratios 
and couplings of both the NMSSM Higgses and SUSY particles for all the surviving data points which have passed 
the various experimental and theoretical constraints implemented in the package NMSSMTools5.1.2 \cite{NMHDECAY1,NMHDECAY2,NMSSMTools}.

\section{\large Higgs boson Signal Rates}
\label{sect:rates}

In order to investigate the discovery prospects for the $h_2$ at the LHC\footnote{We assume a center-of-mass 
energy $\sqrt s=14$ TeV for the LHC.}, we calculate the cross sections for $h_2$
production for the surviving data points by using 
CalcHEP \cite{CalcHEP} . We focus here on the following  production channels: \\
(i) gluon fusion $gg\to h_2$ \\
(ii) associated production with a pair of bottom quarks $g g\to b\bar b h_2$ \\
(iii) associated production with a pair of top quarks $gg\to t\bar t h_2$.\footnote{The production modes
$q\bar q \to b\bar b h_2$ and $q\bar q \to t\bar t h_2$ are negligible at the LHC with $\sqrt{s} = 14$ TeV.}

\begin{figure}
 \centering\begin{tabular}{cc}

 \includegraphics[scale=0.80]{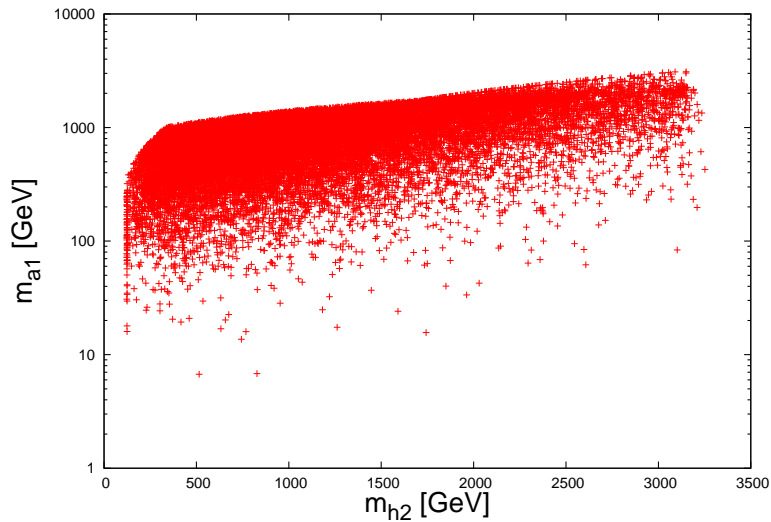}
 \end{tabular}
\caption{The correlations between the second lightest CP-even Higgs mass, $m_{h_2}$
and the lightest CP-odd Higgs masses,  $m_{a_1}$.} 
   
\label{fig0}
\end{figure}

Fig.~\ref{fig0} displays the correlation between the lightest CP-odd Higgs boson mass $m_{a_1}$ 
and the second lightest CP-even Higgs boson mass $m_{h_2}$. Is is shown that the $a_1$ 
can be very light, $\sim 7$ GeV.\footnote{For the mass region $m_{a_1} < 2m_b$,  the decay into $\tau^+\tau^-$ 
is dominant compared to the decay into $c\bar c$ or $\mu^+\mu^-$.} It is also interesting to note that 
the smaller $m_{h_2}$, the smaller $m_{a_1}$. Furthermore, it is
obvious that there are some points of the NMSSM parameter space where the two Higgs states $h_2$ and $a_1$
can simultaneously have the same mass around 125 GeV. 

\begin{figure}
 \centering\begin{tabular}{ccc}
    \includegraphics[scale=0.40]{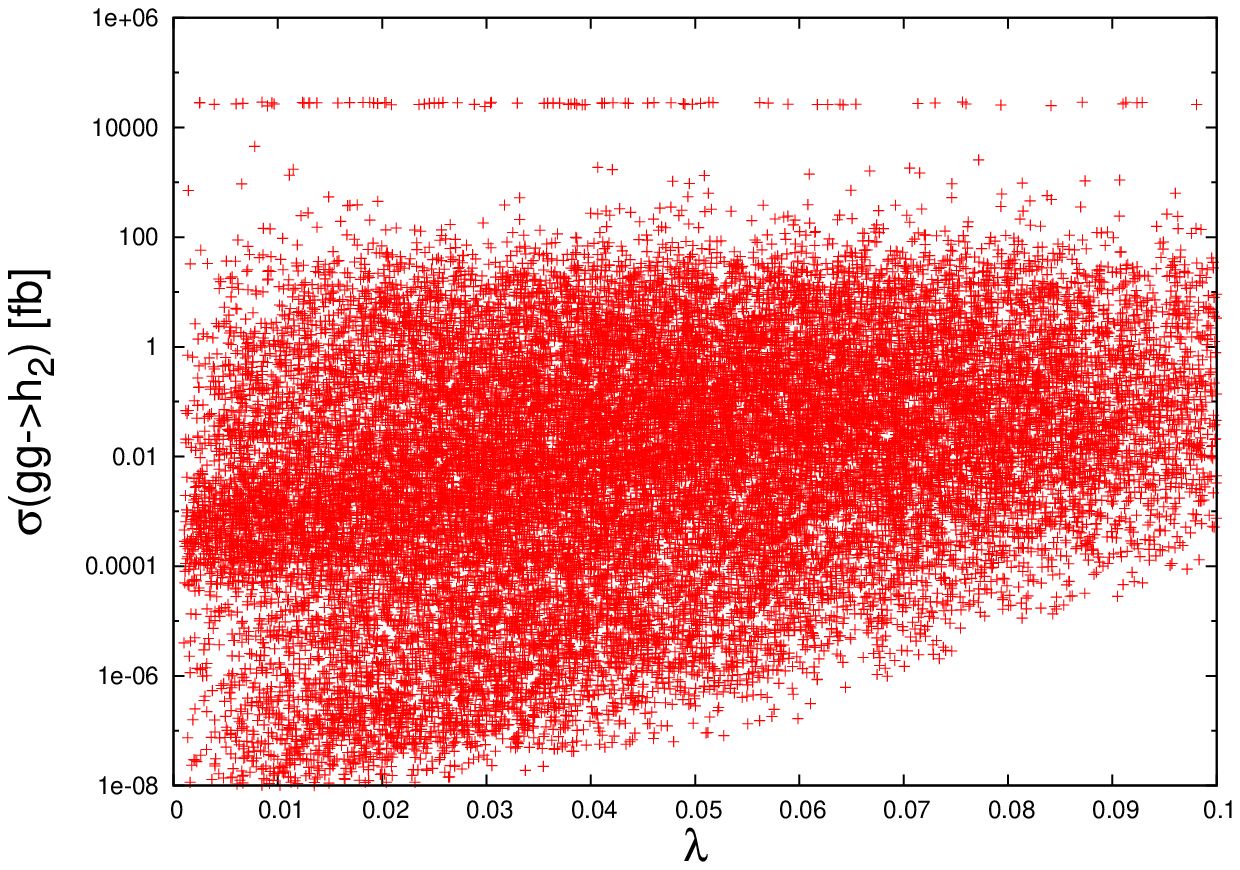}&\includegraphics[scale=0.40]{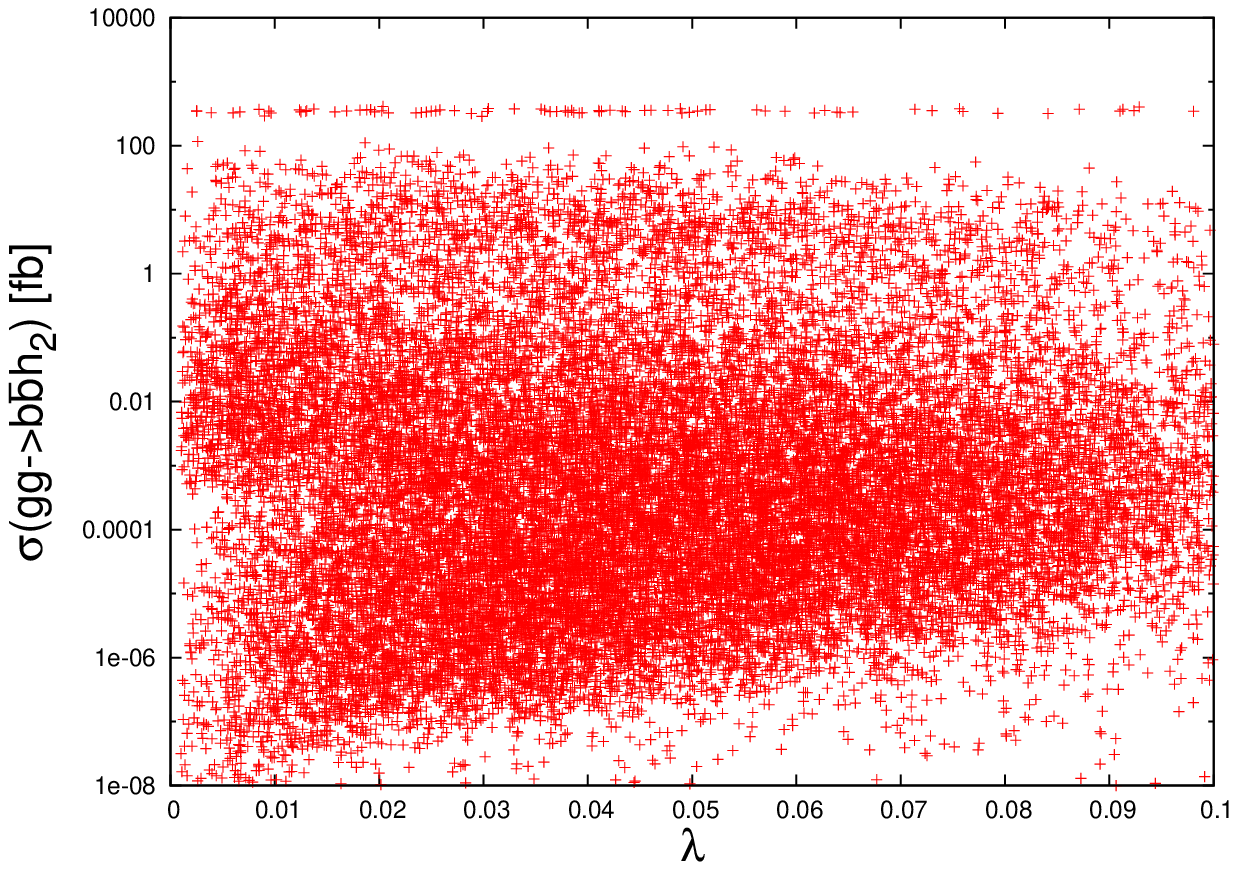}&\includegraphics[scale=0.40]{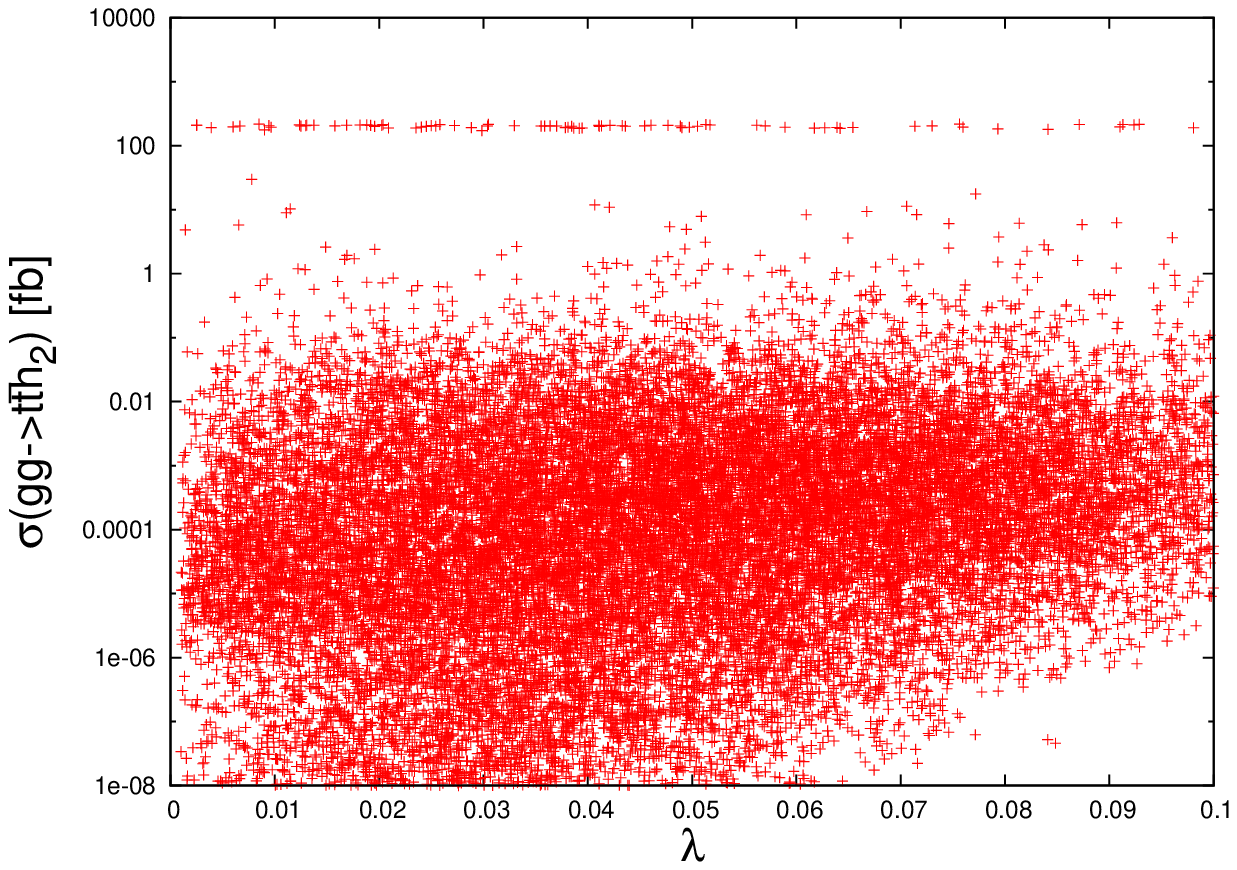}\\
  \includegraphics[scale=0.40]{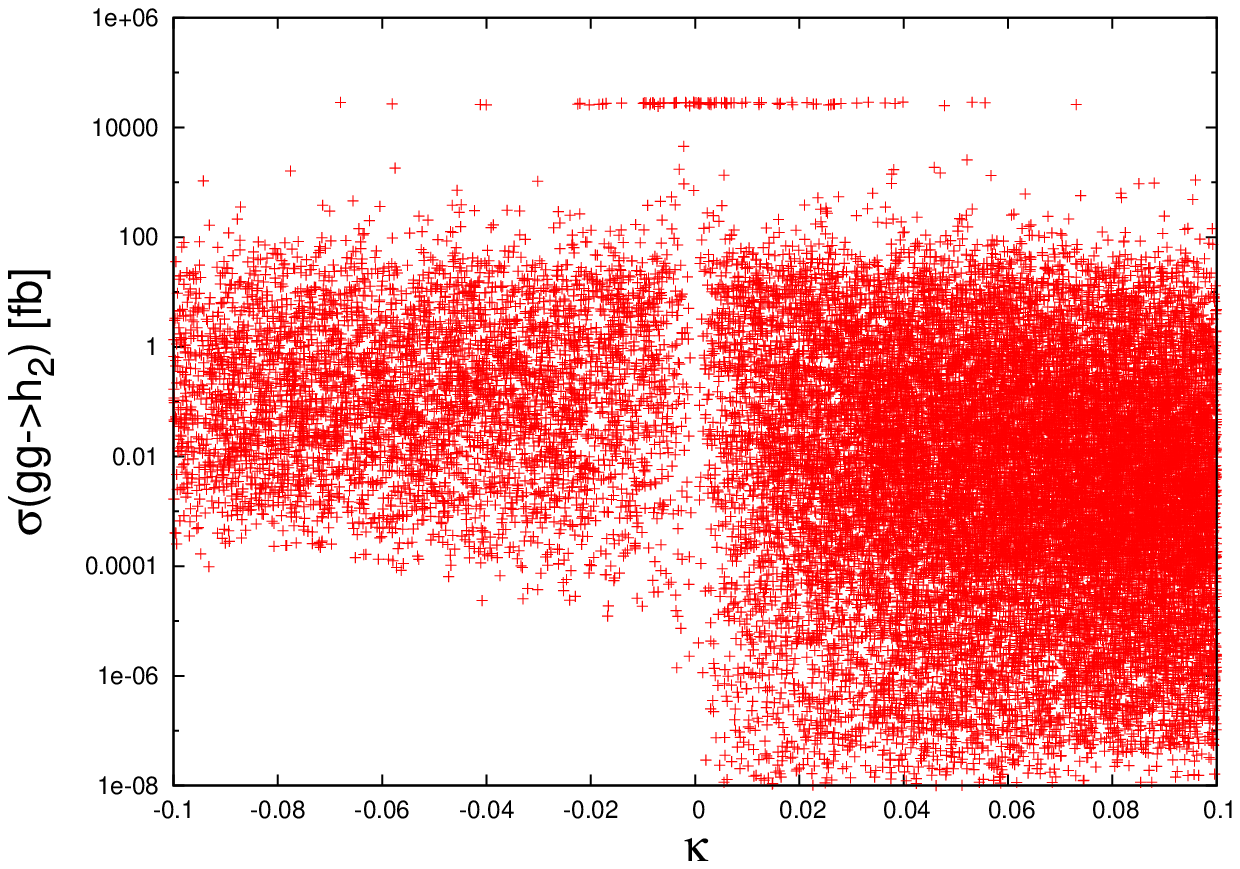}&\includegraphics[scale=0.40]{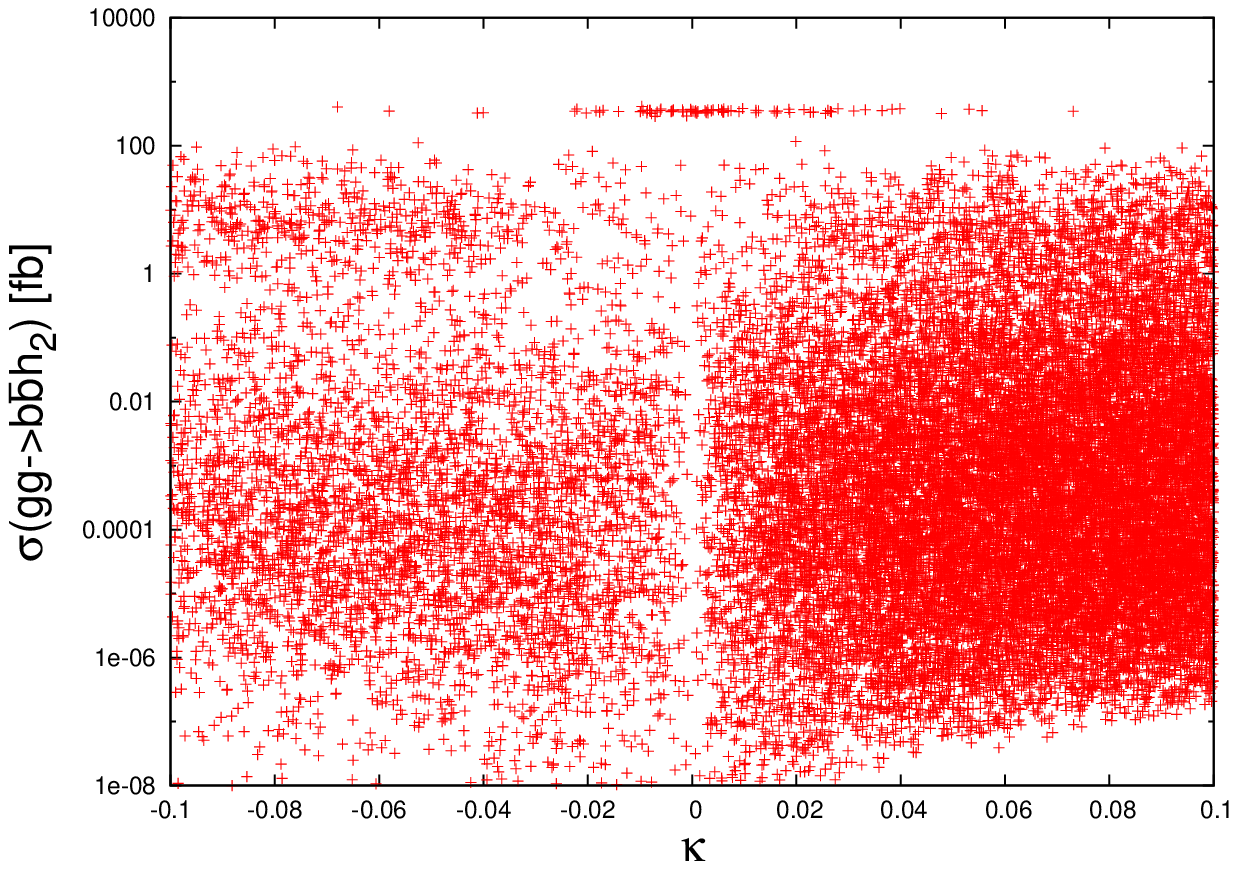}&\includegraphics[scale=0.40]{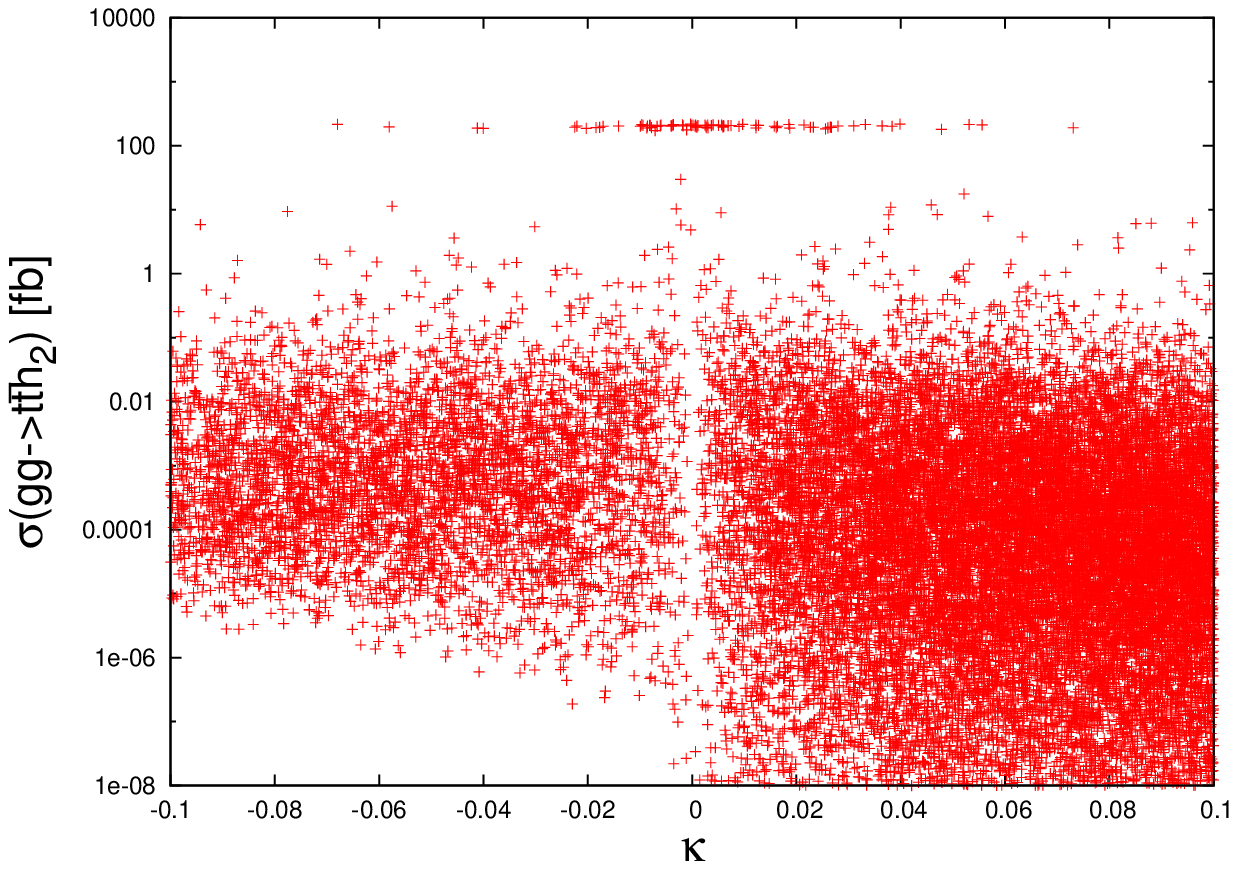}
    \end{tabular}
  
\caption{The cross section for the second lightest CP-even Higgs boson $h_2$ produced in gluon fusion $\sigma(gg\to h_2)$ (left),
in association with bottom quarks $\sigma(gg\to b\bar b h_2)$ (middel) and in association with top quarks $\sigma(gg\to t\bar t h_2)$ (right)
as functions of $\lambda$ (top) and $\kappa$ (bottom).} 

 \label{fig1}
\end{figure}

As an initial step of our analysis, we show in Fig.~\ref{fig1} the cross sections for the 
$h_2$ produced in gluon fusion (left),
in association with bottom quarks (middle) and in association with top quarks (right) against $\lambda$ (top) and $\kappa$ (bottom).
It is clear that in our chosen parameter space the positive values of $\kappa$ are favored,
while the distributions in $\lambda$ are nearly uniform. Also, it is clear that the maximum values of cross sections do not only occur
in the region with nearly vanishing $\lambda$ and $\kappa$, MSSM-limit\footnote{The MSSM-limit can be obtained by simultaneously taking
$\lambda \rightarrow 0$ and $\kappa \rightarrow 0$ while keeping the parameters $\mu_{\rm eff}$, $A_\lambda$ and $A_\kappa$ fixed, which is
not applied in our study},
but also in the other regions because a combination of all the six tree
level parameters jointly affects $m_{h_2}$, which in turn affects the values of cross sections. The maximum cross sections occur when
the $h_2$ becomes more doublet-like than the singlet-like. 

As a second step, we calculate the total cross section times branching ratios of the $h_2$ production at the LHC
in femtobarn (fb): 

\begin{eqnarray*}
\sigma (gg\to h_2) \times {\rm Br}(h_2\to XX), \\
\sigma (gg\to b\bar b h_2) \times {\rm Br}(h_2\to XX), \\
\sigma (gg\to t\bar t h_2) \times {\rm Br}(h_2\to XX),
\end{eqnarray*}

where $h_2\to XX$  represents the $h_2$ decay channels with $XX$ = $\tau^+\tau^-$, $b\bar b$,
$t\bar t$, $\gamma\gamma$, $Z\gamma$, $W^+W^-$, $ZZ$, $a_1a_1$, $h_1h_1$ and $Za_1$.

\begin{figure}
 \centering\begin{tabular}{ccc}
    \includegraphics[scale=0.40]{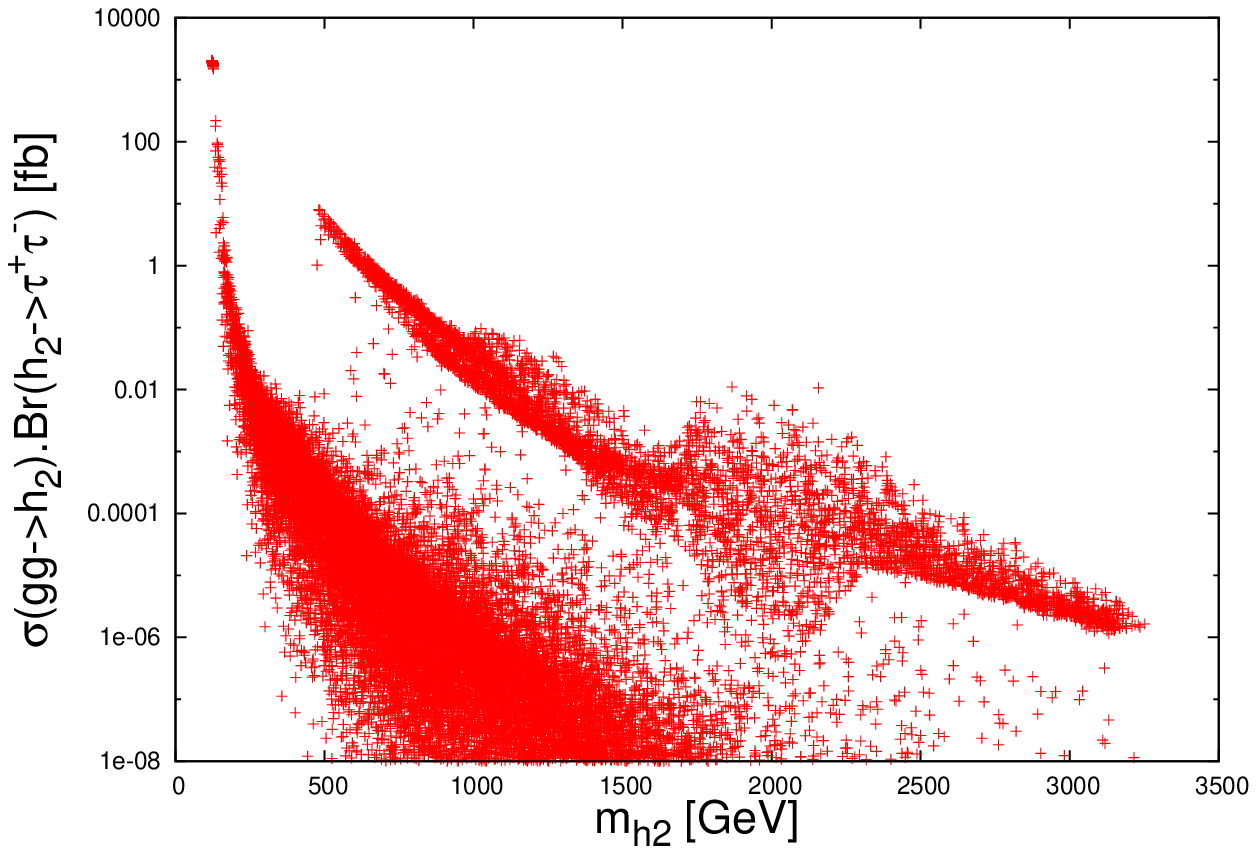}&\includegraphics[scale=0.40]{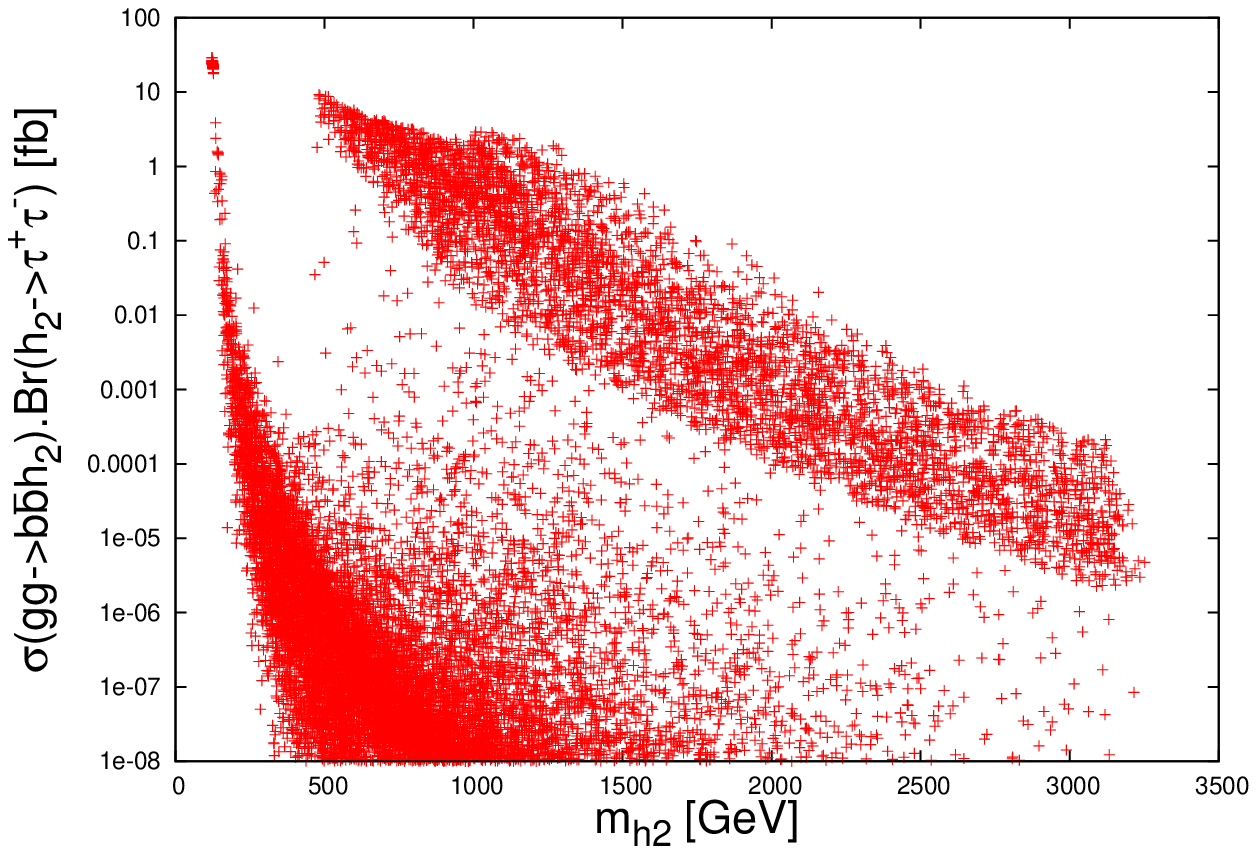}&\includegraphics[scale=0.40]{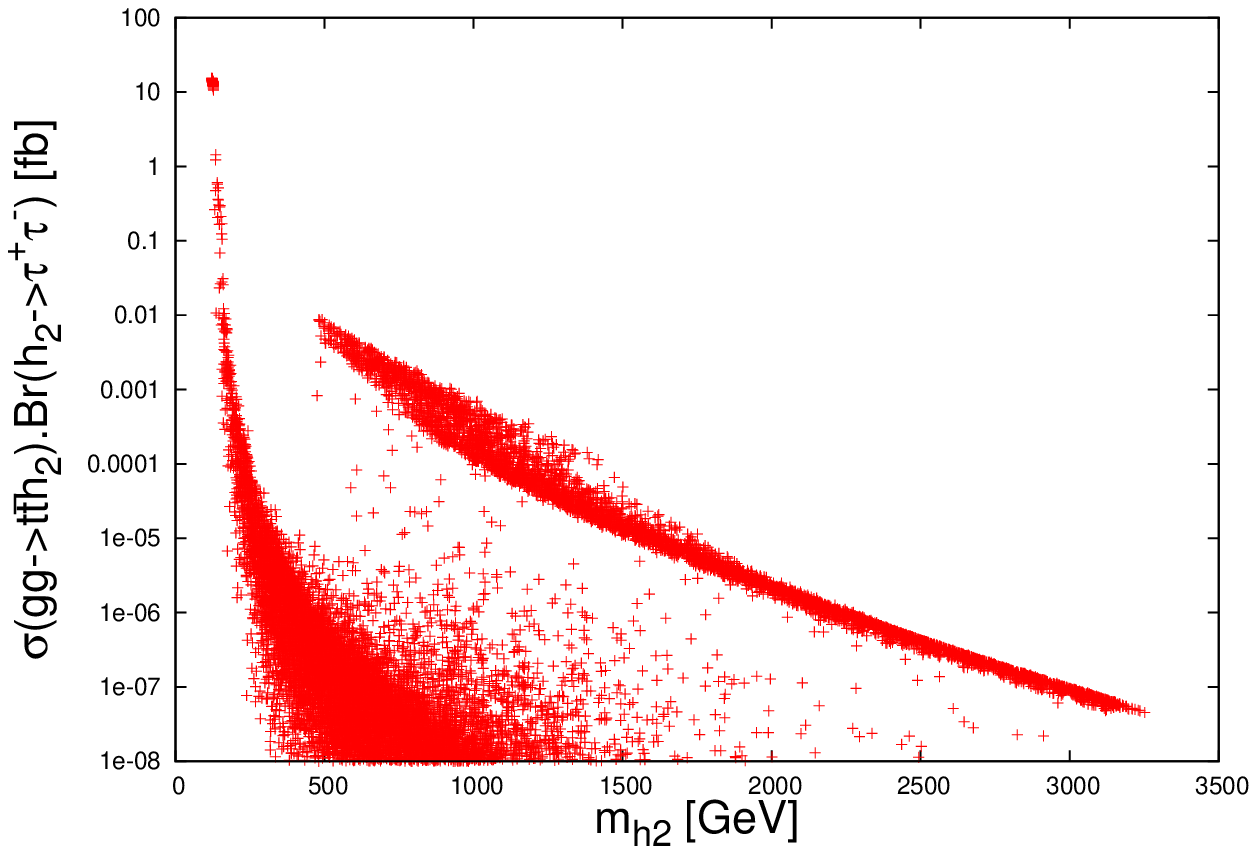}\\
  \includegraphics[scale=0.40]{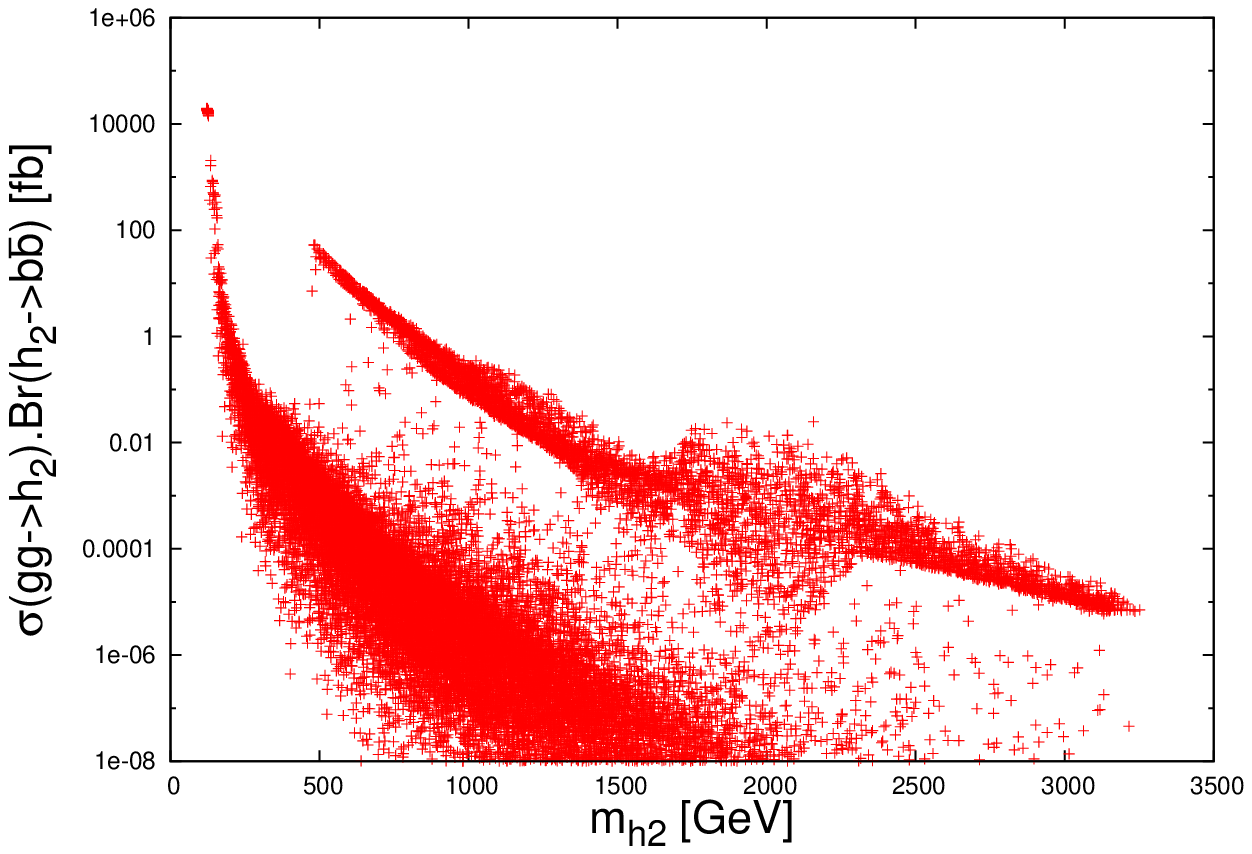}&\includegraphics[scale=0.40]{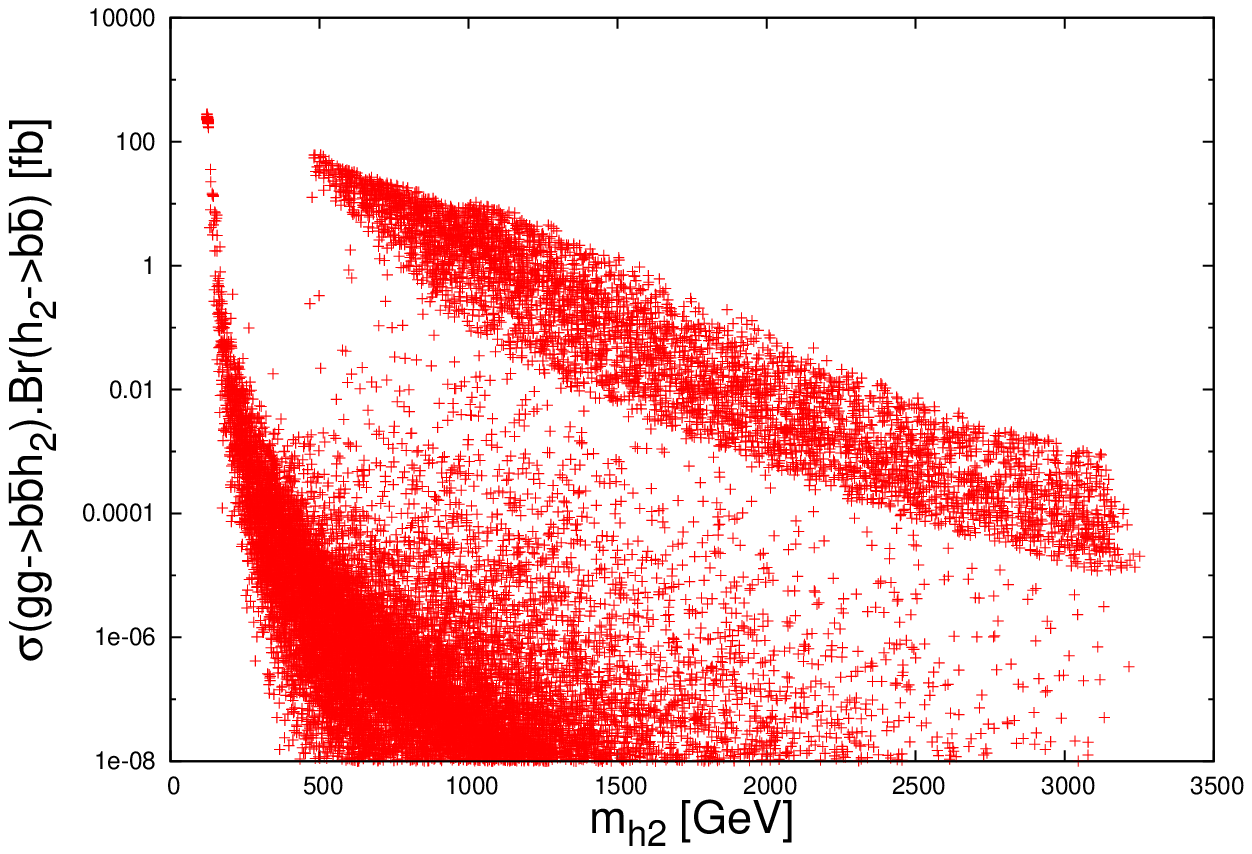}&\includegraphics[scale=0.40]{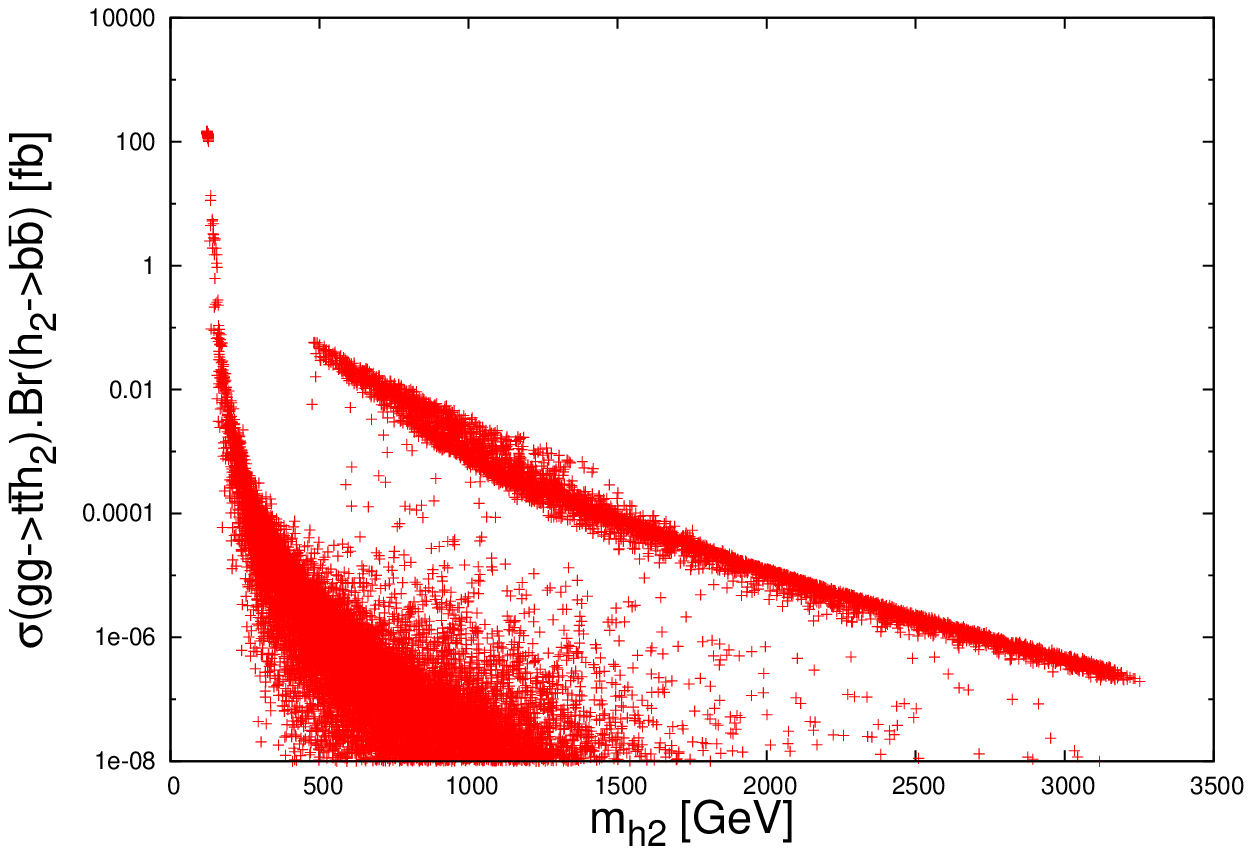}\\
  \includegraphics[scale=0.40]{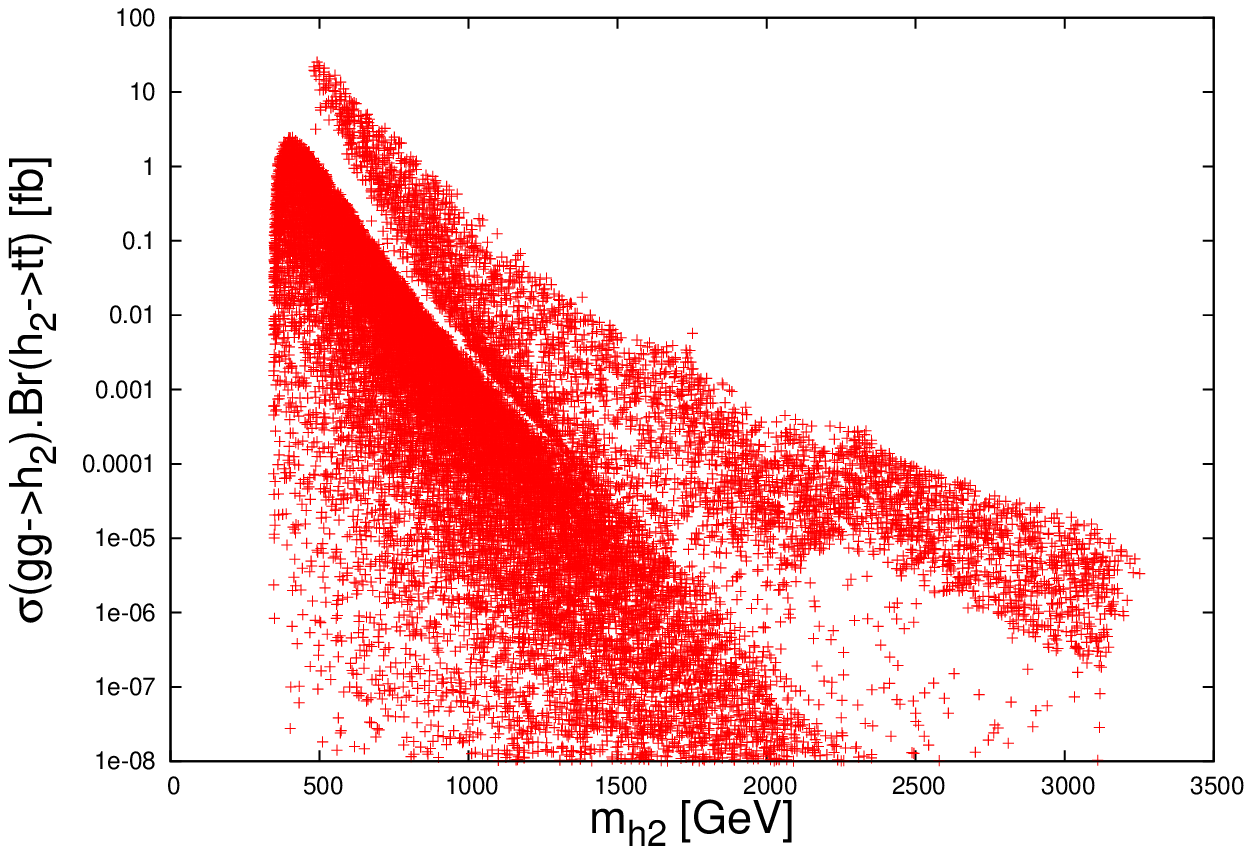}&\includegraphics[scale=0.40]{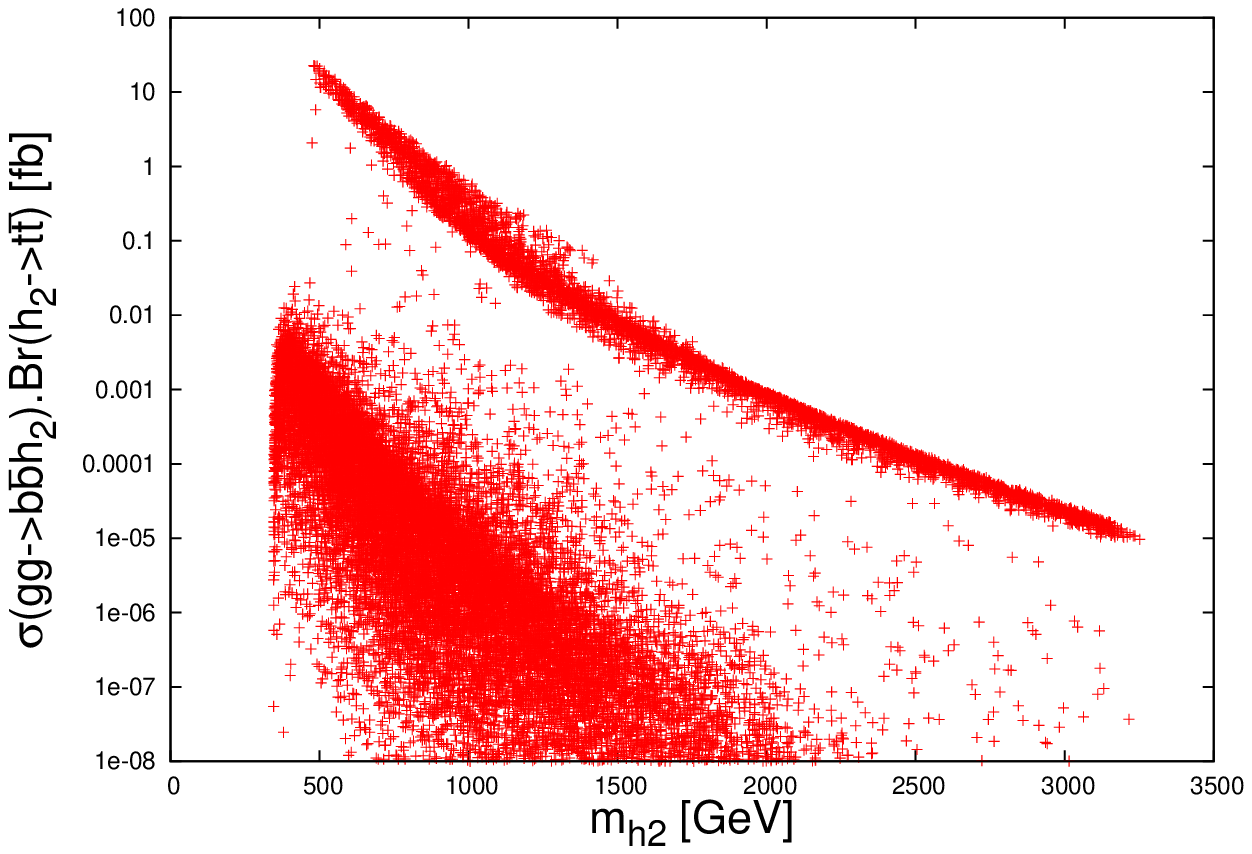}&\includegraphics[scale=0.40]{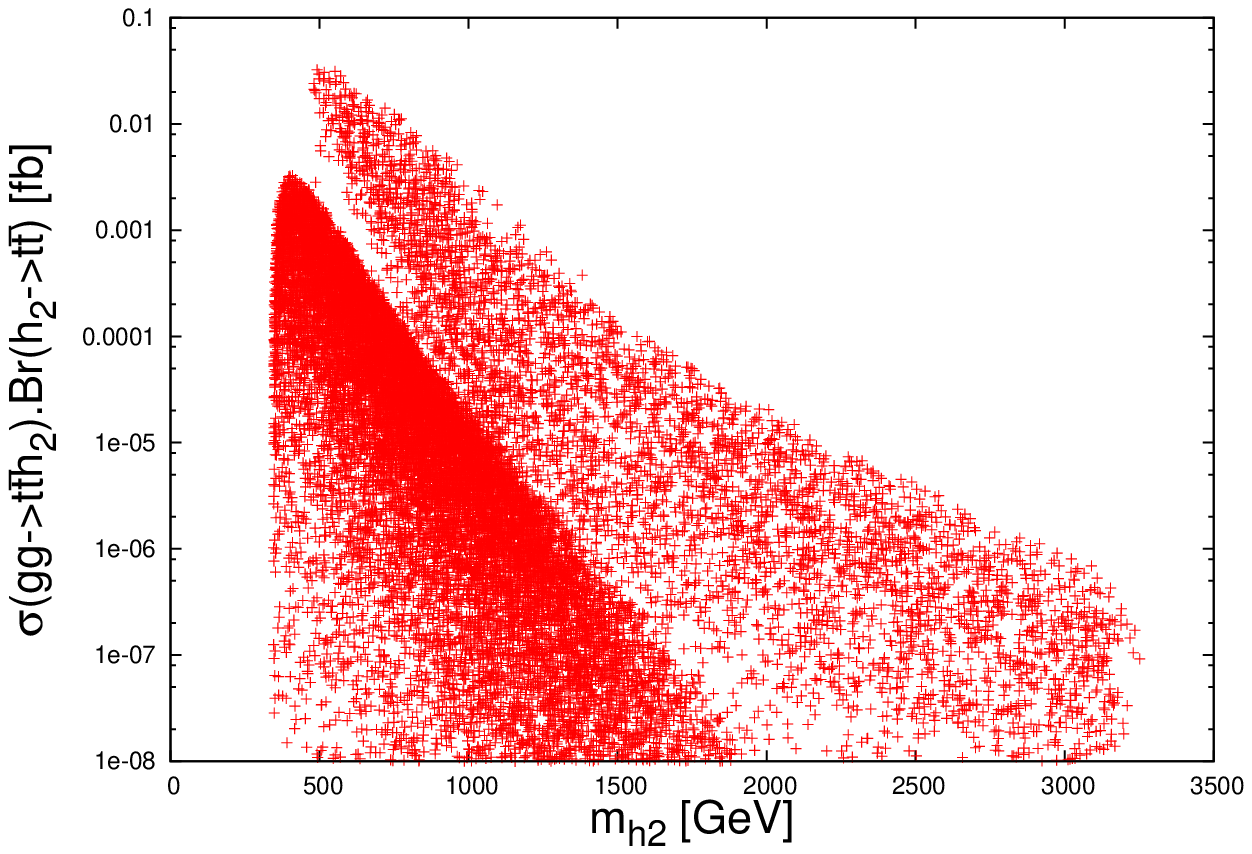}
   \end{tabular}
  
\caption{The signal rates for the second lightest CP-even Higgs boson $h_2$ produced in gluon fusion $\sigma(gg\to h_2)$ (left),
in association with bottom quarks $\sigma(gg\to b\bar b h_2)$ (middel) and in association with top quarks $\sigma(gg\to t\bar t h_2)$ (right)
times $~{\rm Br}(h_2\to \tau^+\tau^-)$ (top), $~{\rm Br}(h_2\to b\bar b)$ (middle) and $~{\rm Br}(h_2\to t\bar t)$ (bottom) as functions of $m_{h_2}$.}

 \label{fig2}
\end{figure}

Figure ~\ref{fig2} displays the signal rates for the $h_2$ produced in gluon fusion (left),
in association with bottom quarks (middle) and in association with top quarks (right) 
for $\tau^+\tau^-$ (top), $b\bar b$ (middle) and $t\bar t$ (bottom) final states as functions of $m_{h_2}$. 
We note that there is a gap in the figure with only few points between the upper region and lower region of the the parameter space
in the mass range $m_{h_2} \gtrsim 480$ GeV. This is because 
the decay $h_2\to a_1a_1$, which is dominant in large area of the NMSSM parameter space, is closed or suppressed in the upper region.
It is clear from the figure that the cross section times branching ratio 
for the gluon fusion process is the highest one in general,
reaching up to 2600 fb, 26000 fb and 
60 fb for $\tau^+\tau^-$, $b\bar b$ and $t\bar t$ final states, respectively, see the left panels of the figure.
The inclusive cross section for the production in association with bottom quarks is the second largest one, 
topping 30 fb, 300 fb and 25 fb in case of $\tau^+\tau^-$, $b\bar b$ and $t\bar t$ final states, respectively, see the middle panels of
the figure.\footnote{Searches at the LHC for additional neutral MSSM Higgs bosons set some limits for the production of such bosons.
 For $\tau$ lepton final states, these  range  from  18000 fb at 90 GeV to 3.5 fb at 3.2 TeV  for
production via gluon fusion and from 15000 fb at 90 GeV to 2.5 fb at 3.2 TeV for production in
association with b quarks at a center-of-mass energy of 13 TeV, corresponding to an integrated luminosity  of  35.9 $fb^{-1}$,
 see Ref. ~\refcite{Sirunyan:2018zut}. For b quark final states, 
the upper limits on Higgs boson produced in
association with at least one additional b quark  range from about
250000 fb at 100 GeV  to about 1000 fb at 900 GeV at a center-of-mass energy of
8 TeV at the LHC, corresponding to an integrated luminosity of 19.7 $fb^{-1}$, see Ref. ~\refcite{Khachatryan:2015tra}. }
However, there is some
region of the NMSSM parameter space in the mass range $m_{h_2} \gtrsim 480$ GeV where
the production in association with bottom quarks can give the largest
signal rate. This is because the $h_2$ in this region has a large $H_d$ component, which can be substantial for large values of tan$\beta$,
so its couplings to down-type fermions are enhanced compared to the SM. In fact, the $h_2$ production in association with bottom quarks
has an extra advantage where 
the associated $b\bar b$ pair can be tagged, allowing a useful handle
for background rejection. 
The production in association with a top-quark pair has only a considerable signal rates for the low mass range $m_{h_2} \lesssim 2 m_{W}$, 
giving maximum rates of about 15 fb and 150 fb in case of $\tau^+\tau^-$ and $b\bar b$  final states, respectively,
see the top left and middle left panels of the figure.
In short, the size of the signal rates in some regions of the parameter space
is quite large and that could help discovering the $h_2$ in the $\tau^+\tau^-$ and  $b\bar b$ final states. Also, the $t\bar t$ final states
may be exploited to discover the $h_2$ in gluon fusion and in association
with bottom-quark pair for $m_{h_2} \sim 480$ GeV or slightly above, though challenging due to large backgrounds
and a complicated $t\bar t$ final state.

\begin{figure}
 \centering\begin{tabular}{ccc}
  \includegraphics[scale=0.40]{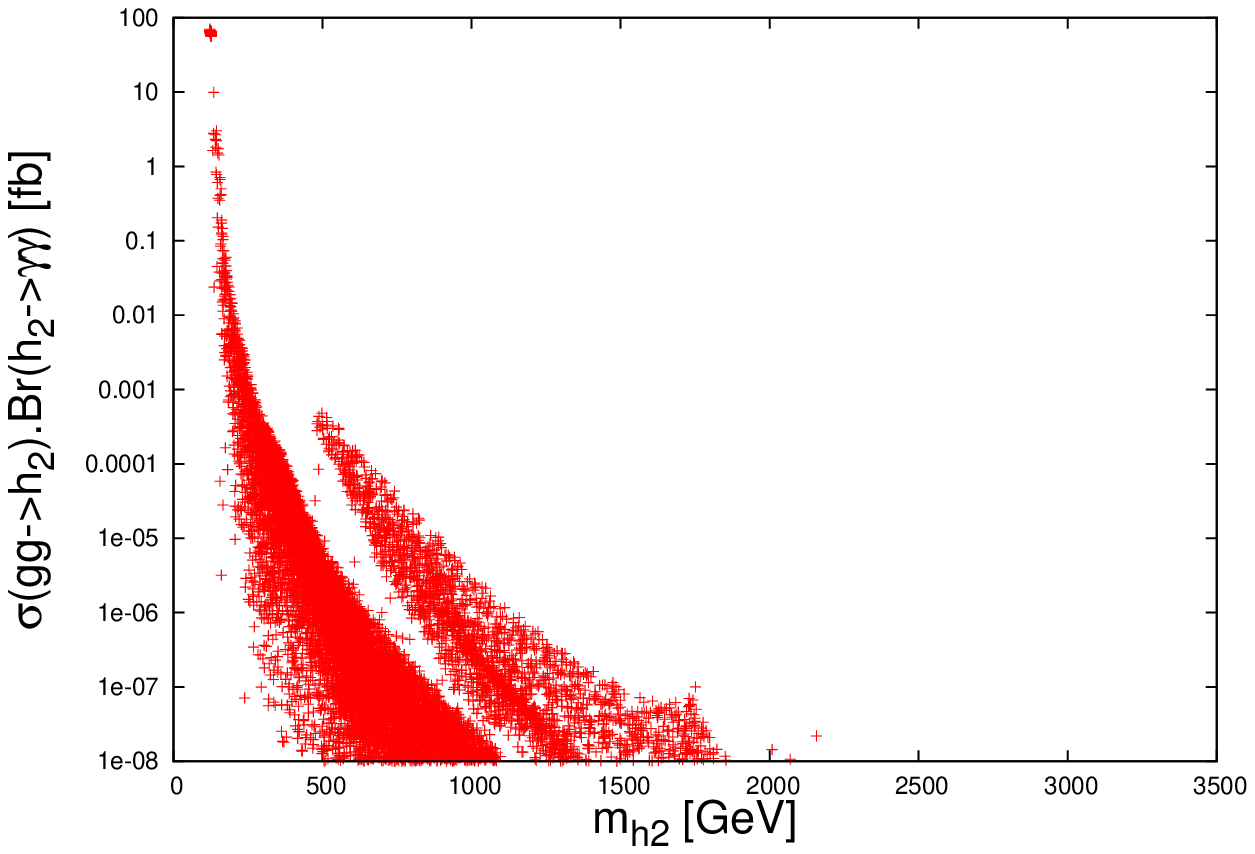}&\includegraphics[scale=0.40]{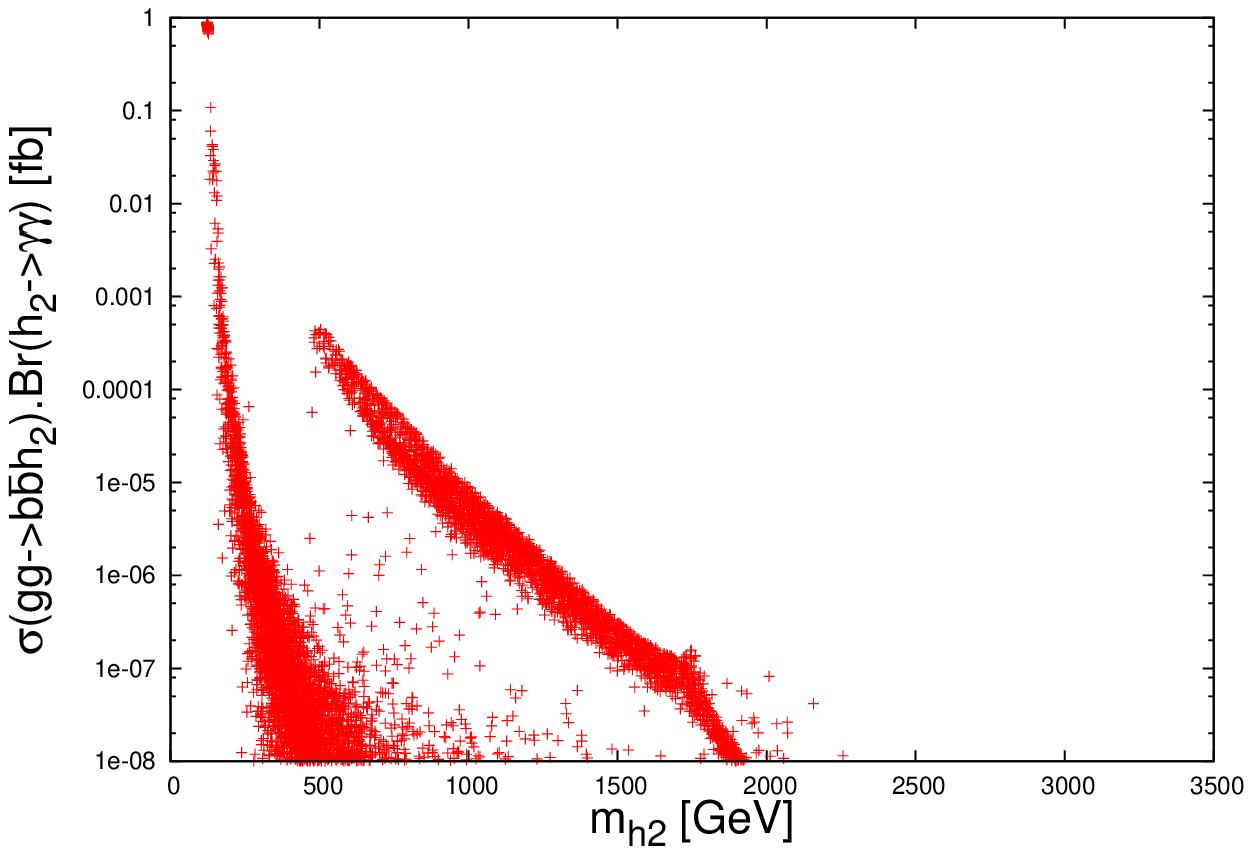}&\includegraphics[scale=0.40]{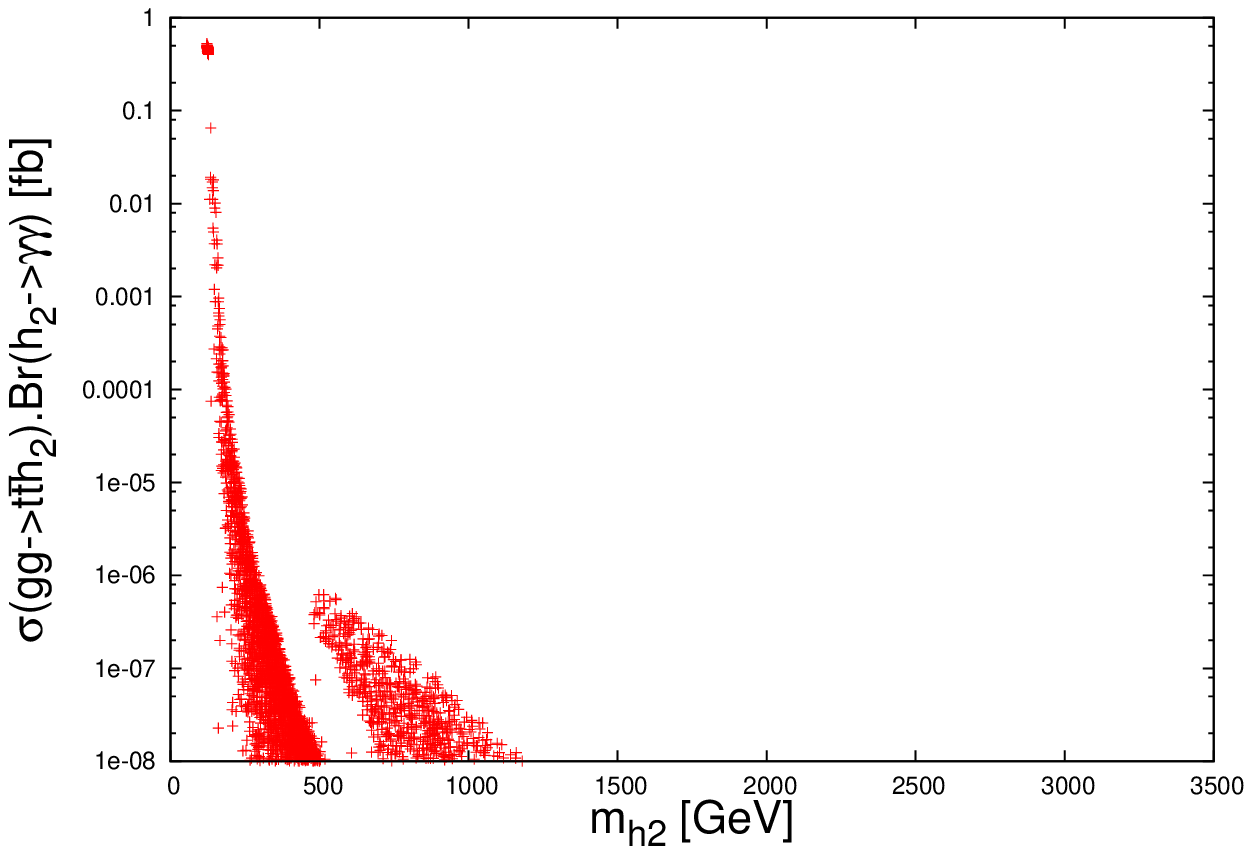}\\
    \includegraphics[scale=0.40]{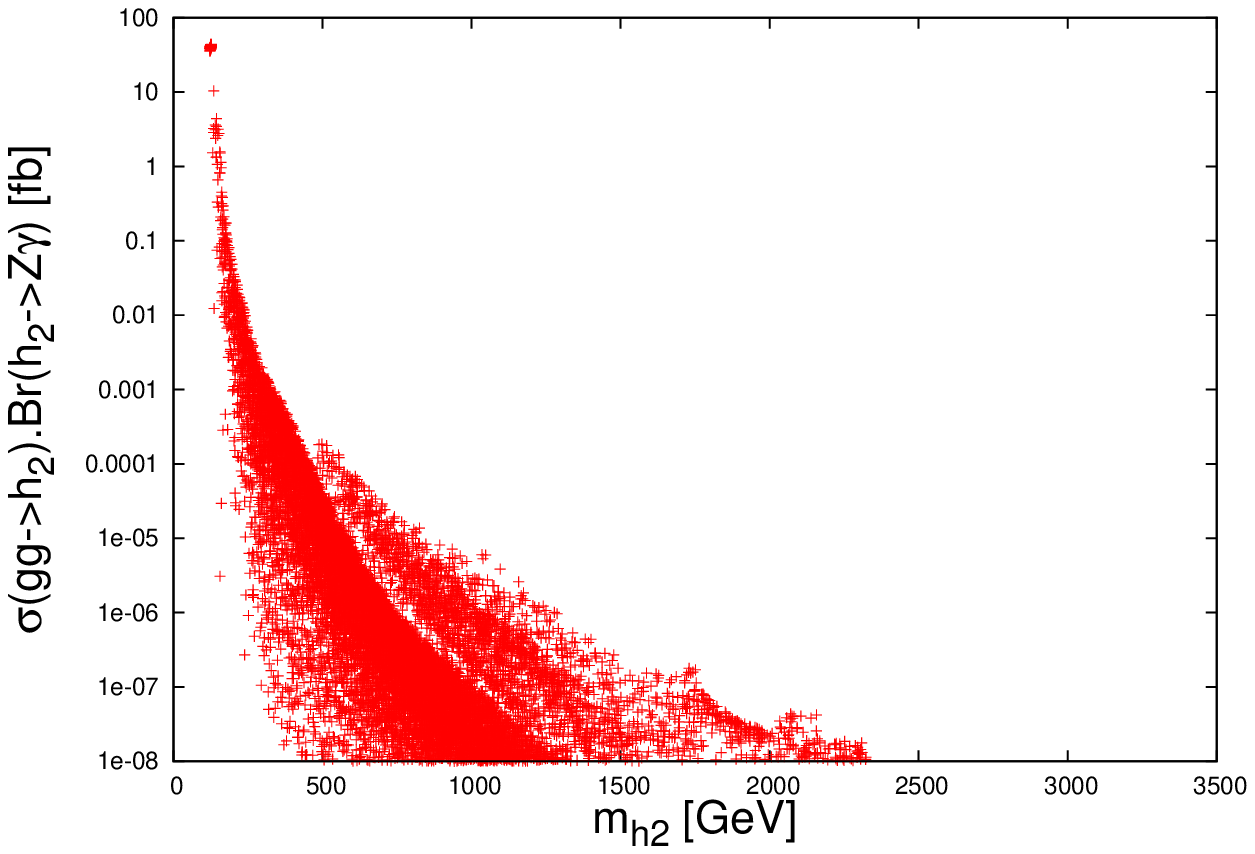}&\includegraphics[scale=0.40]{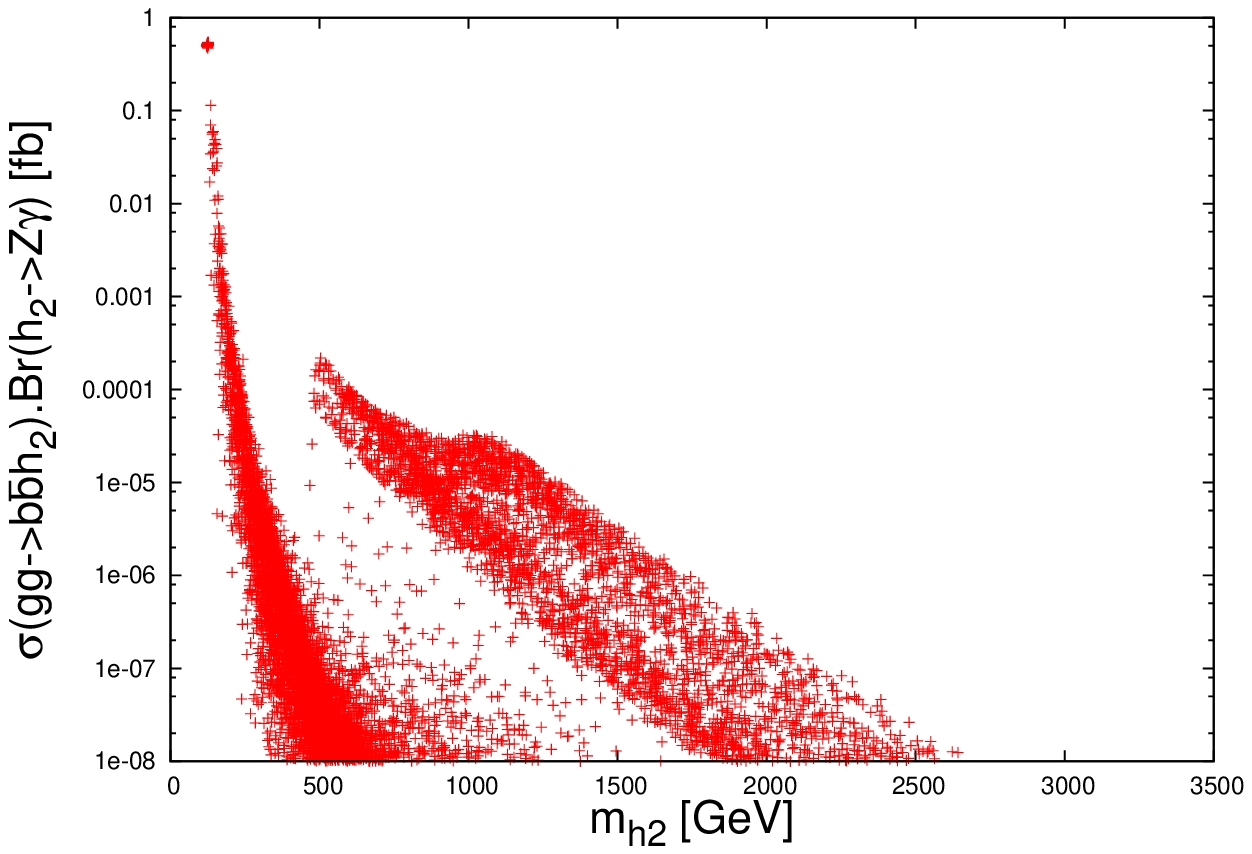}&\includegraphics[scale=0.40]{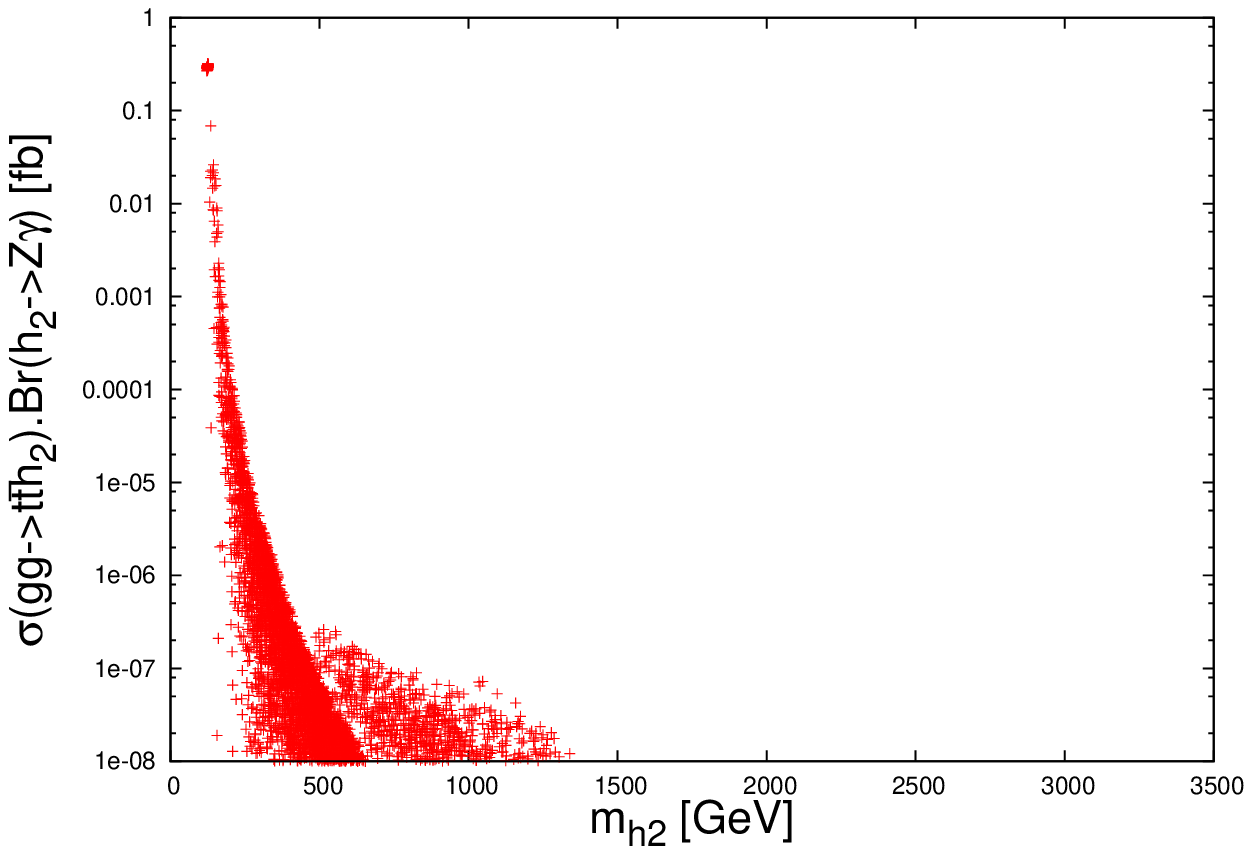}
   
   \end{tabular}

\caption{The signal rates for the second lightest CP-even Higgs boson $h_2$ produced in gluon fusion $\sigma(gg\to h_2)$ (left),
in association with bottom quarks $\sigma(gg\to b\bar b h_2)$ (middel) and  in association with top quarks $\sigma(gg\to t\bar t h_2)$ (right)
times $~{\rm Br}(h_2\to \gamma\gamma)$ (top) and $~{\rm Br}(h_2\to Z\gamma)$ (bottom) as functions of $m_{h_2}$.}

\label{fig3}
\end{figure}

\begin{figure}
 \centering\begin{tabular}{ccc}
  \includegraphics[scale=0.40]{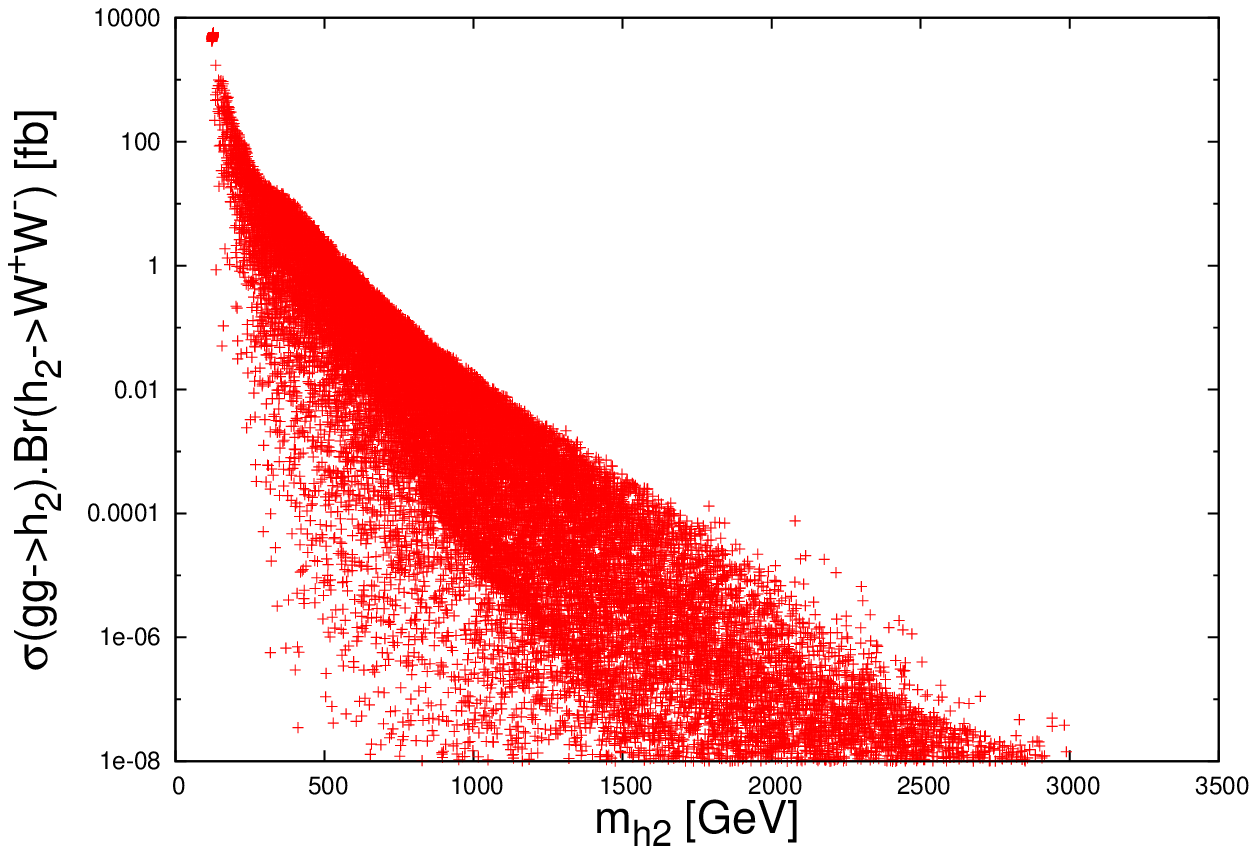}&\includegraphics[scale=0.40]{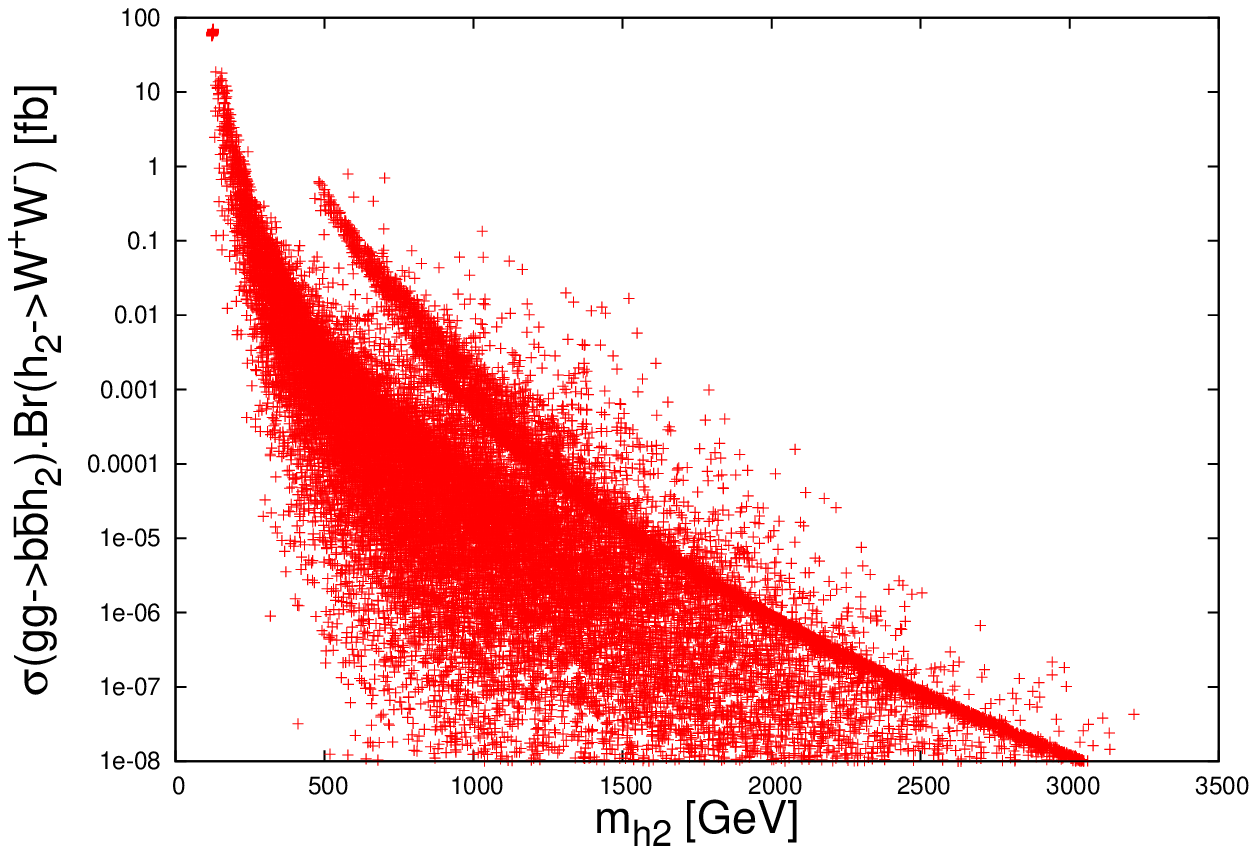}&\includegraphics[scale=0.40]{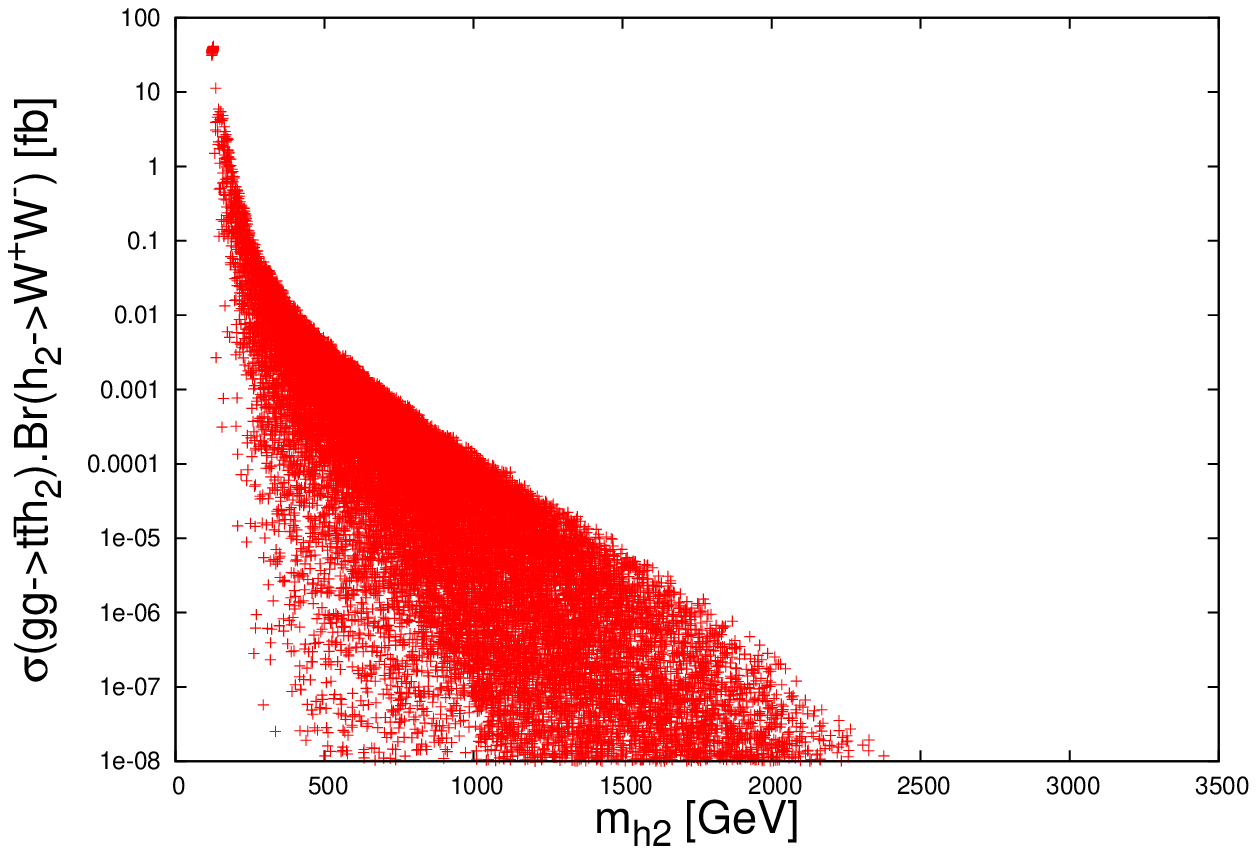}\\
    \includegraphics[scale=0.40]{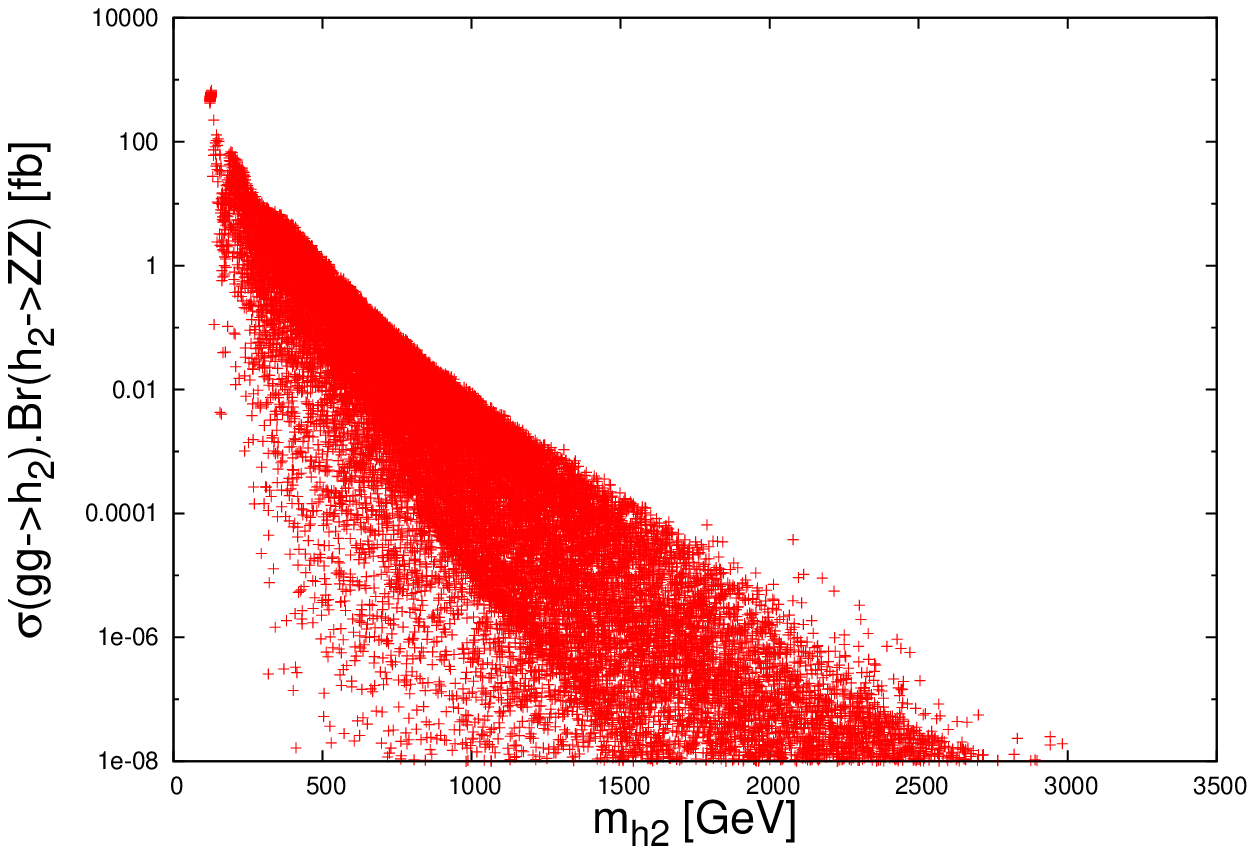}&\includegraphics[scale=0.40]{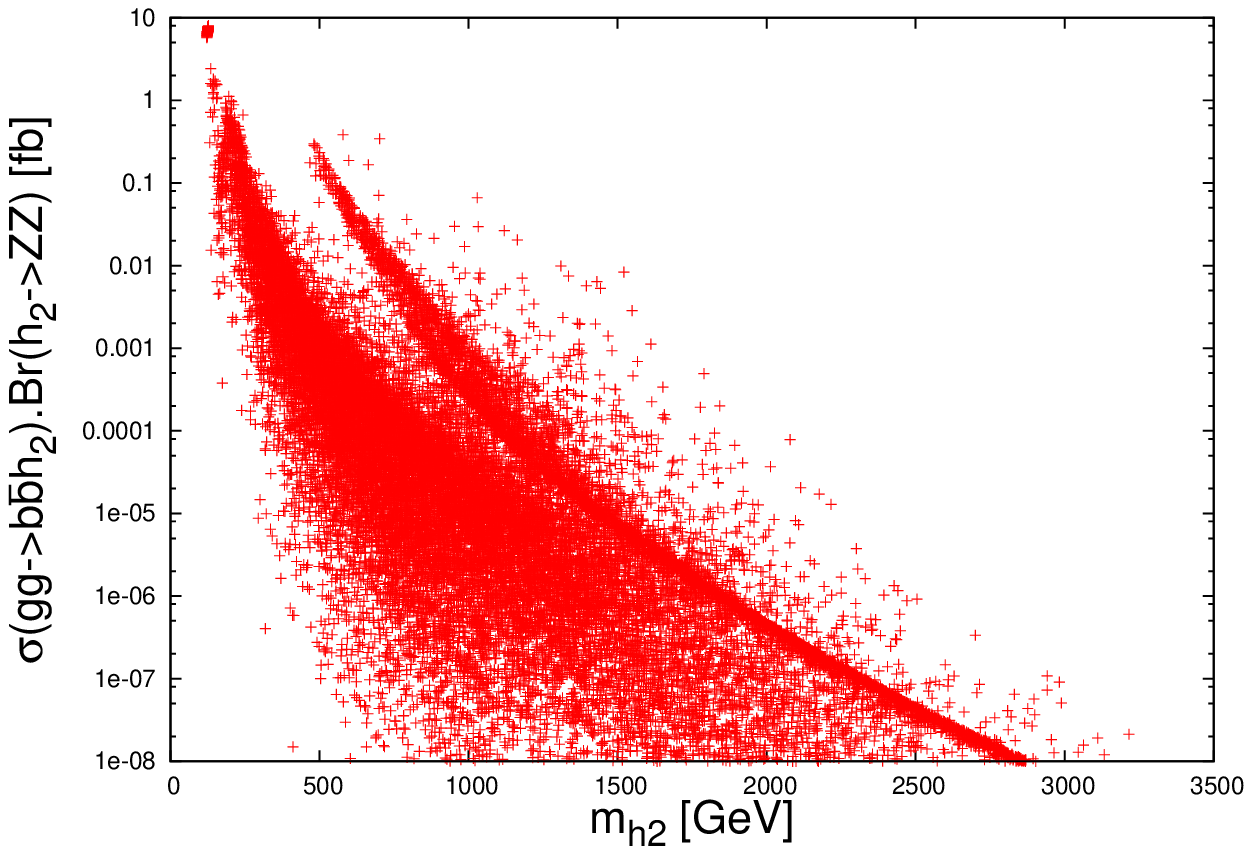}&\includegraphics[scale=0.40]{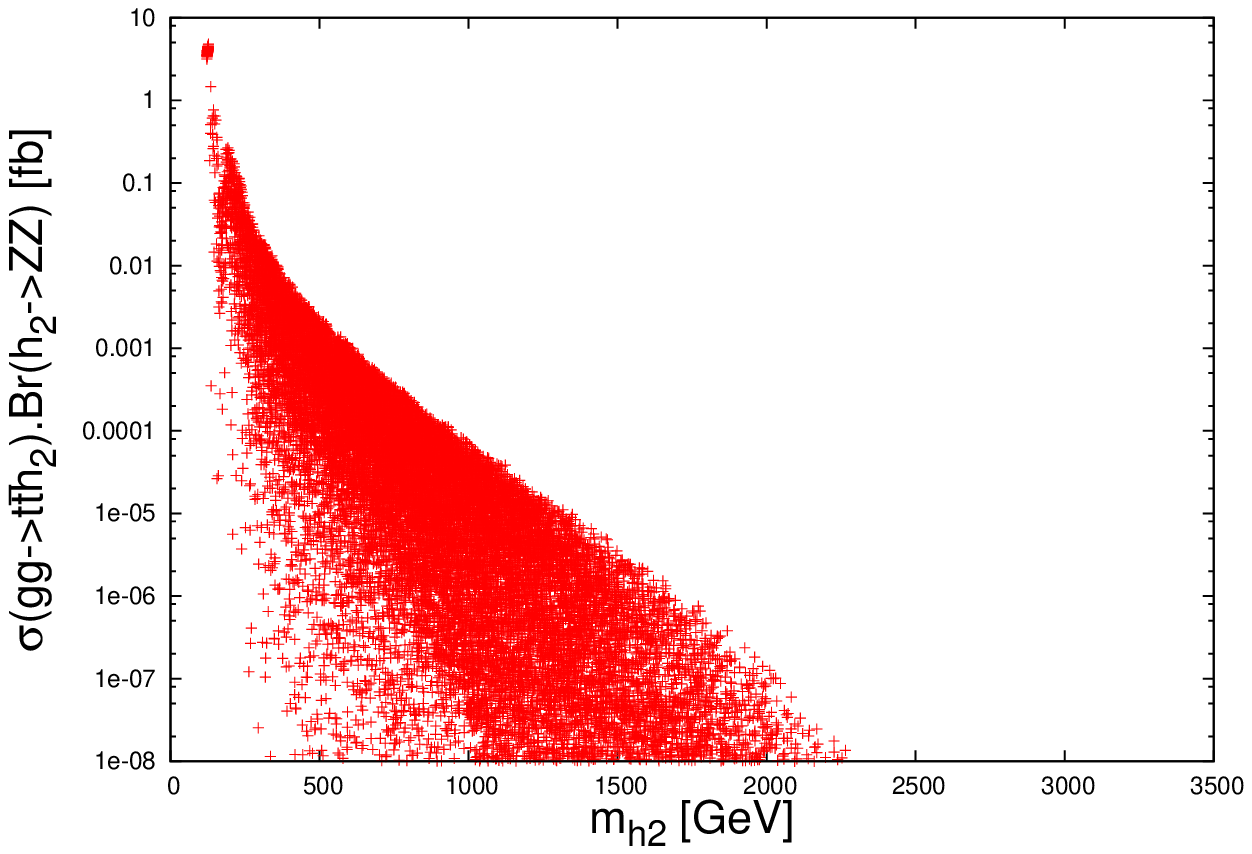}
    \end{tabular}
   
\caption{The signal rates for the second-to-lightest CP-even Higgs boson $h_2$ produced in gluon fusion $\sigma(gg\to h_2)$ (left),
in association with bottom quarks $\sigma(gg\to b\bar b h_2)$ (middel) and in association with top quarks $\sigma(gg\to t\bar t h_2)$ (right)
times $~{\rm Br}(h_2\to W^+W^-)$ (top) and $~{\rm Br}(h_2\to ZZ)$ (bottom) as functions of $m_{h_2}$.}

 \label{fig4}
\end{figure}

In Figs.~\ref{fig3} and \ref{fig4}, we present the production rates for the $h_2$ into the SM bosonic particles. It is shown that
the $\gamma\gamma$ and $Z\gamma$ final states have only sizable signal rates for the low mass range $m_{h_2} \lesssim 2 m_{W}$, see Fig.~\ref{fig3}.
It is clear that the largest production rates comes from gluon fusion channel, reaching maximum 70 fb for the former and 45 fb for the latter,
see the left panels of the figure. The production rates for $h_2$
produced in association with bottom quarks and top quarks are quite small below 1 fb level for both $\gamma\gamma$ and $Z\gamma$ final states. 
Although the signal rates for the $\gamma\gamma$ and $Z\gamma$ final states are not so large but these final states 
offer clean signatures with manageable backgrounds at the LHC, so one may exploit these channels to discover the $h_2$ in the low mass range
especially in gluon fusion production channel. 
Fig.~\ref{fig4} shows the signal rates in the $W^+W^-$ and ZZ final states. It is shown that   
these rates reach up to 10000 fb for the former and 1000 fb for the latter in gluon fusion channel,
see the left panels of the figure.\footnote{The CMS collaboration published results of searches for a heavy neutral Higgs boson decaying
into $WW$ and $ZZ$ bosons with up to 19.7 $fb^{-1}$ of data collected at a centre-of-mass energy of 8 TeV. It was found that
such a heavy Higgs boson with SM-like couplings and decays in the mass range from 145 GeV to 1000 GeV was excluded,
see Ref. ~\refcite{Khachatryan:2015cwa}.} The $h_2$ production in association with both bottom and top quarks has approximately
two orders of magnitude less than that of the gluon fusion production channel, see the middle and right panels.

\begin{figure}
 \centering\begin{tabular}{ccc}
  \includegraphics[scale=0.40]{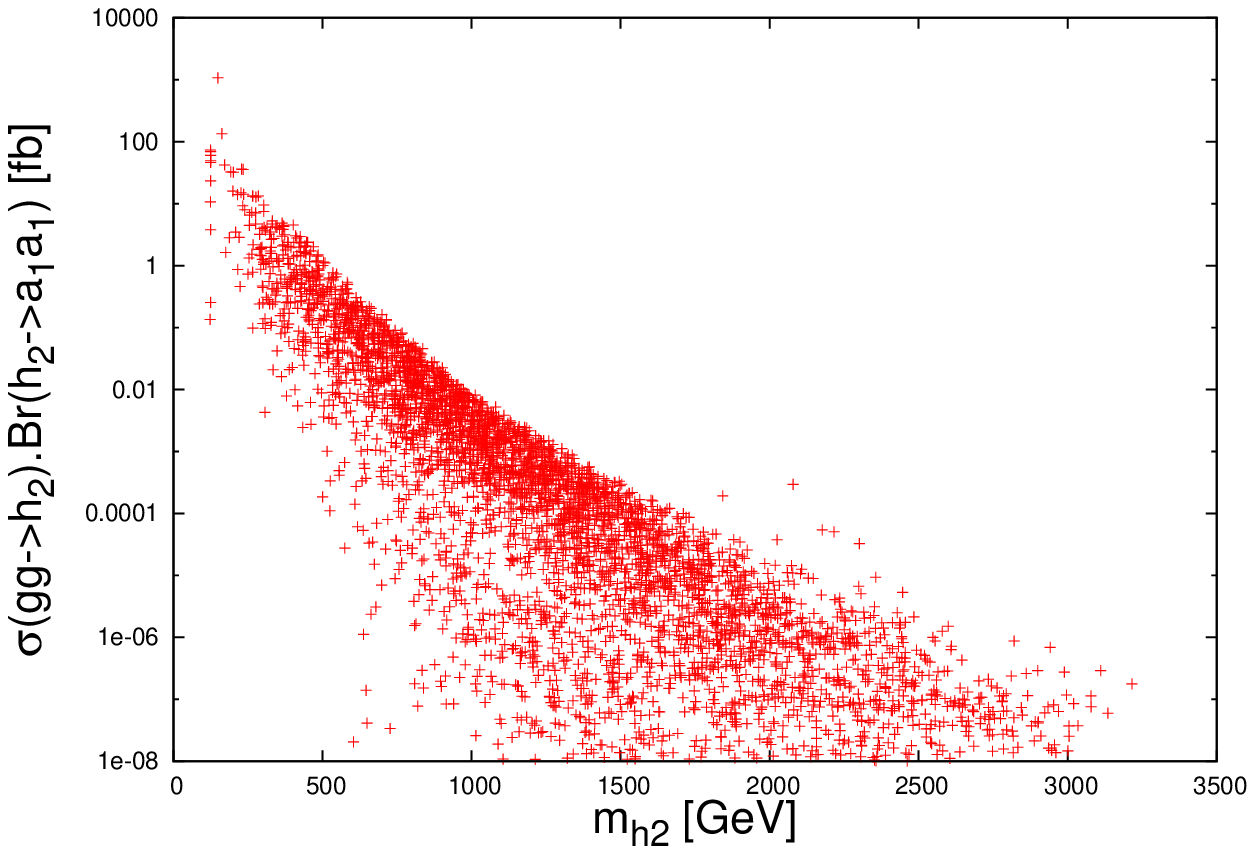}&\includegraphics[scale=0.40]{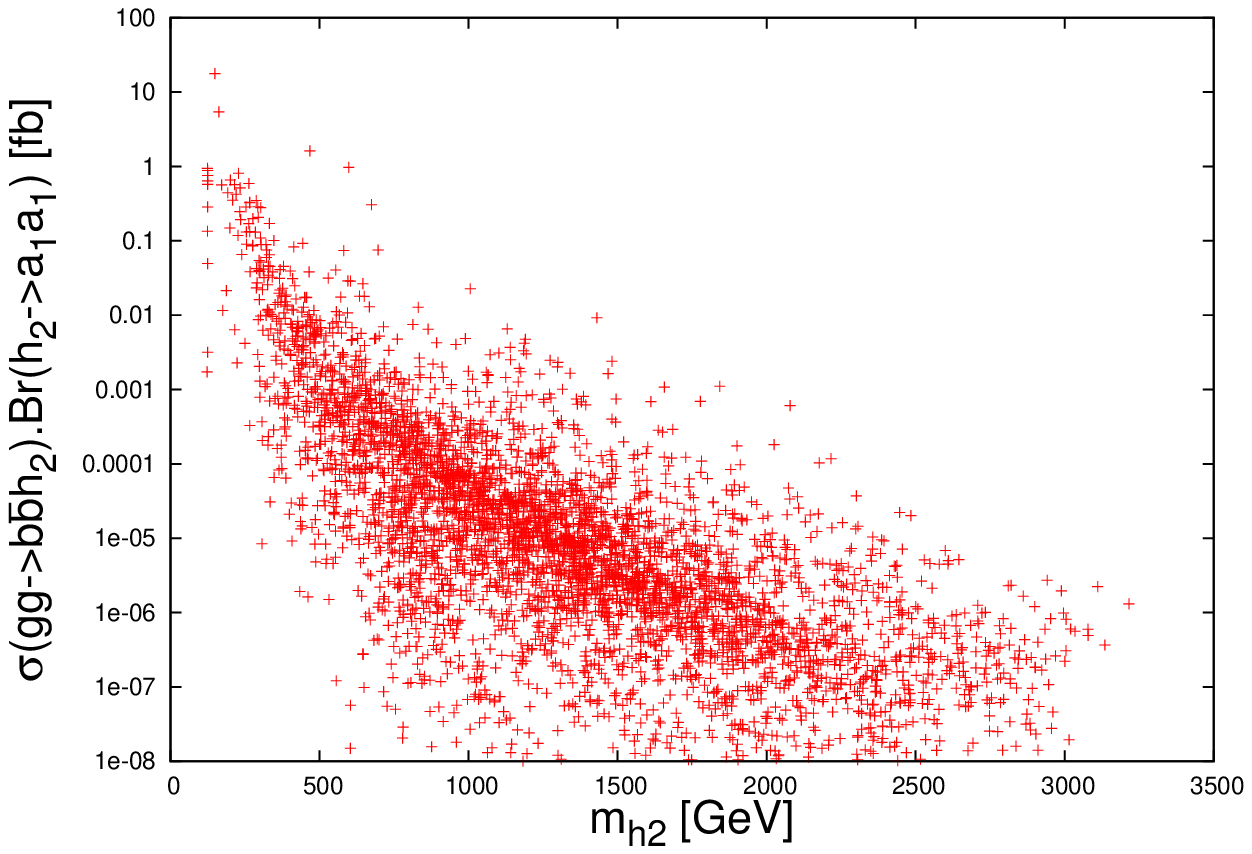}&\includegraphics[scale=0.40]{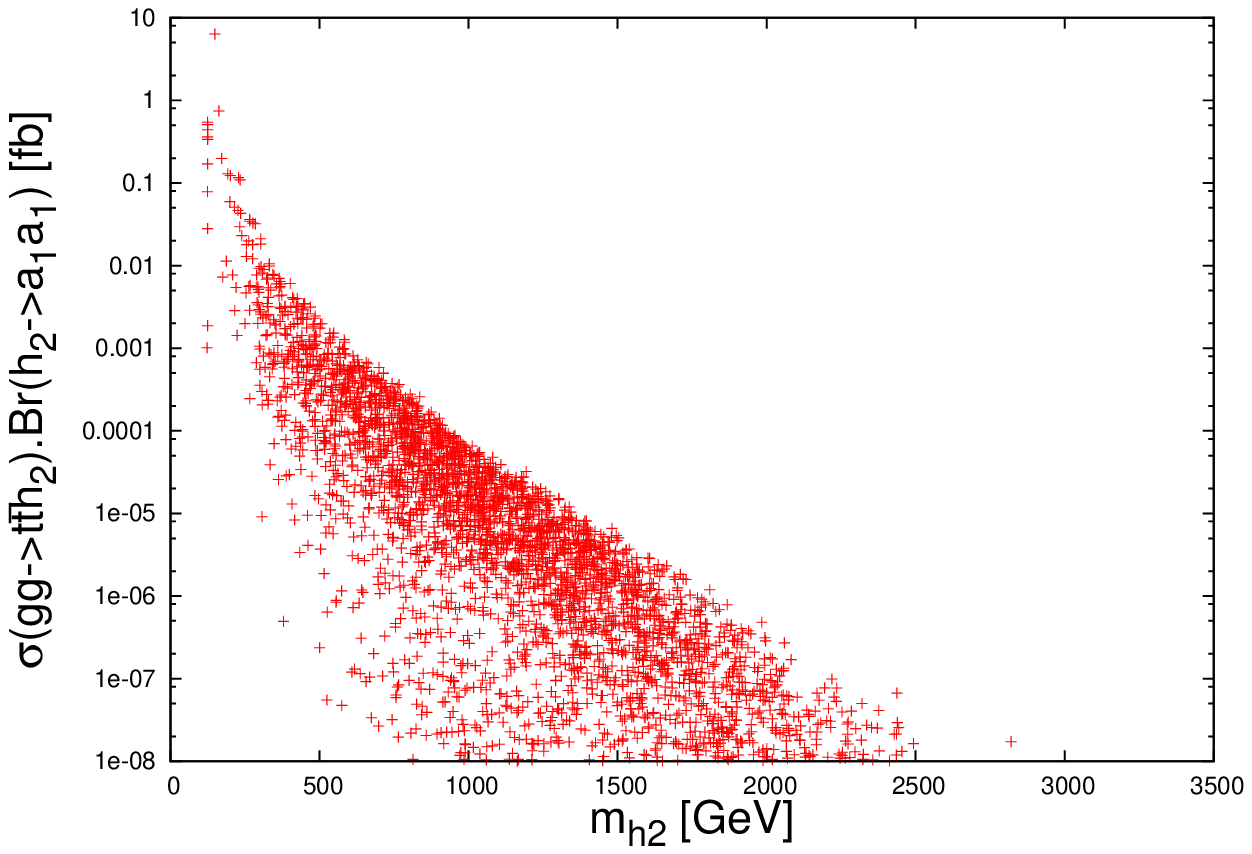}\\
    \includegraphics[scale=0.40]{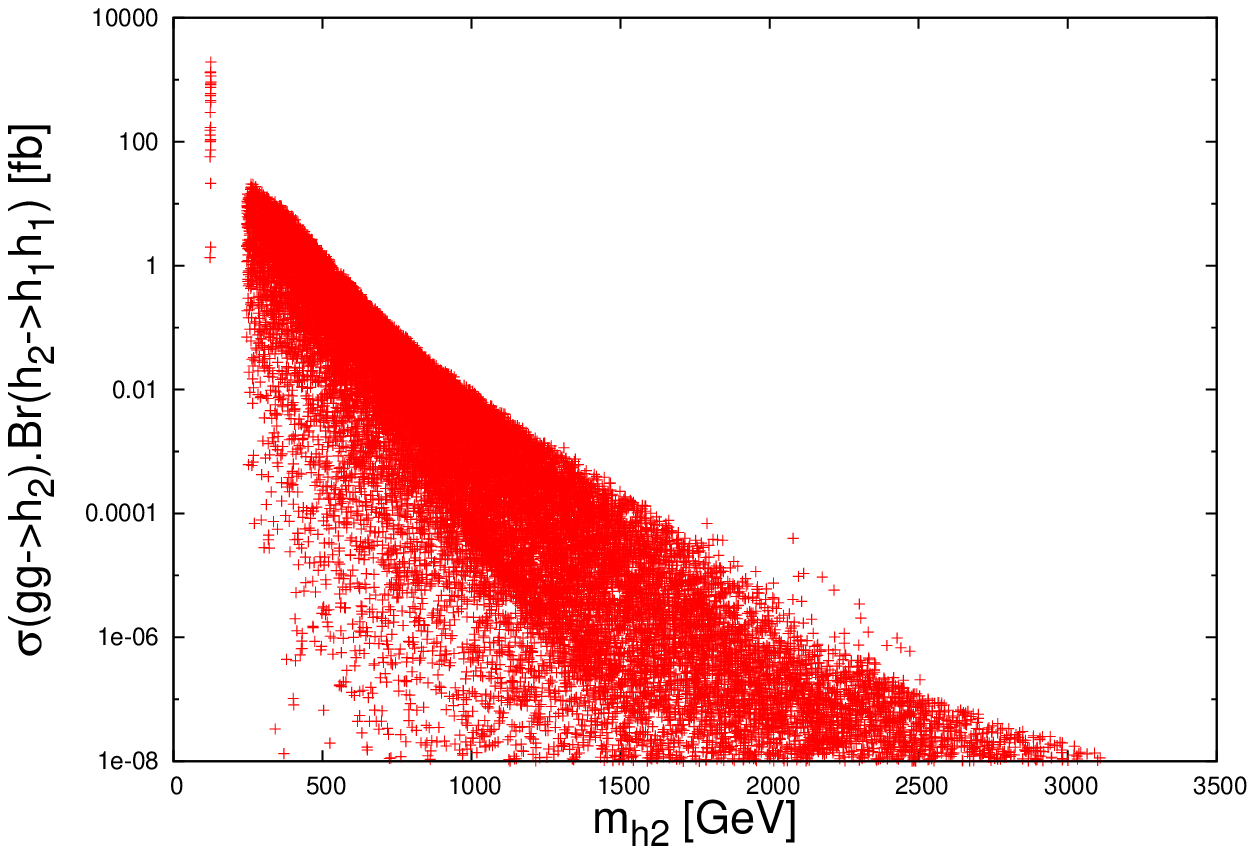}&\includegraphics[scale=0.40]{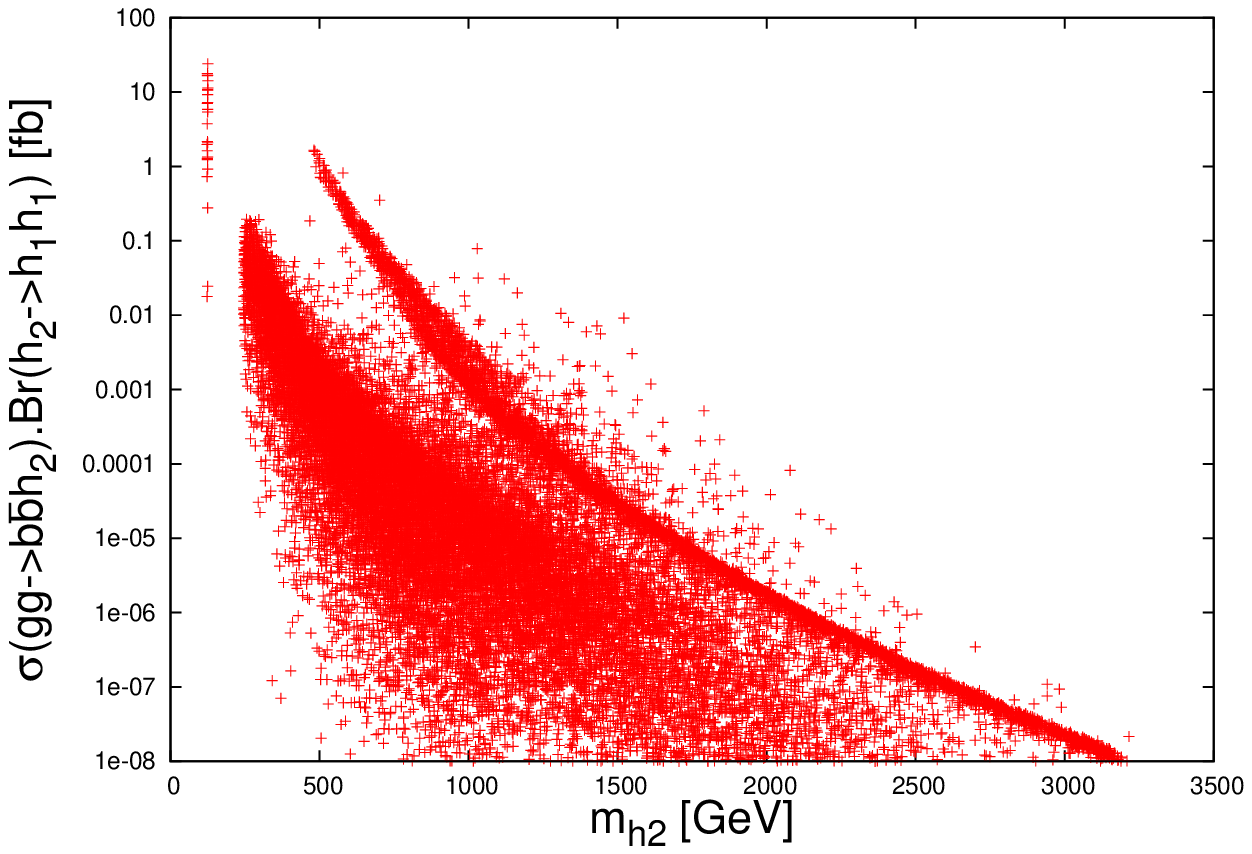}&\includegraphics[scale=0.40]{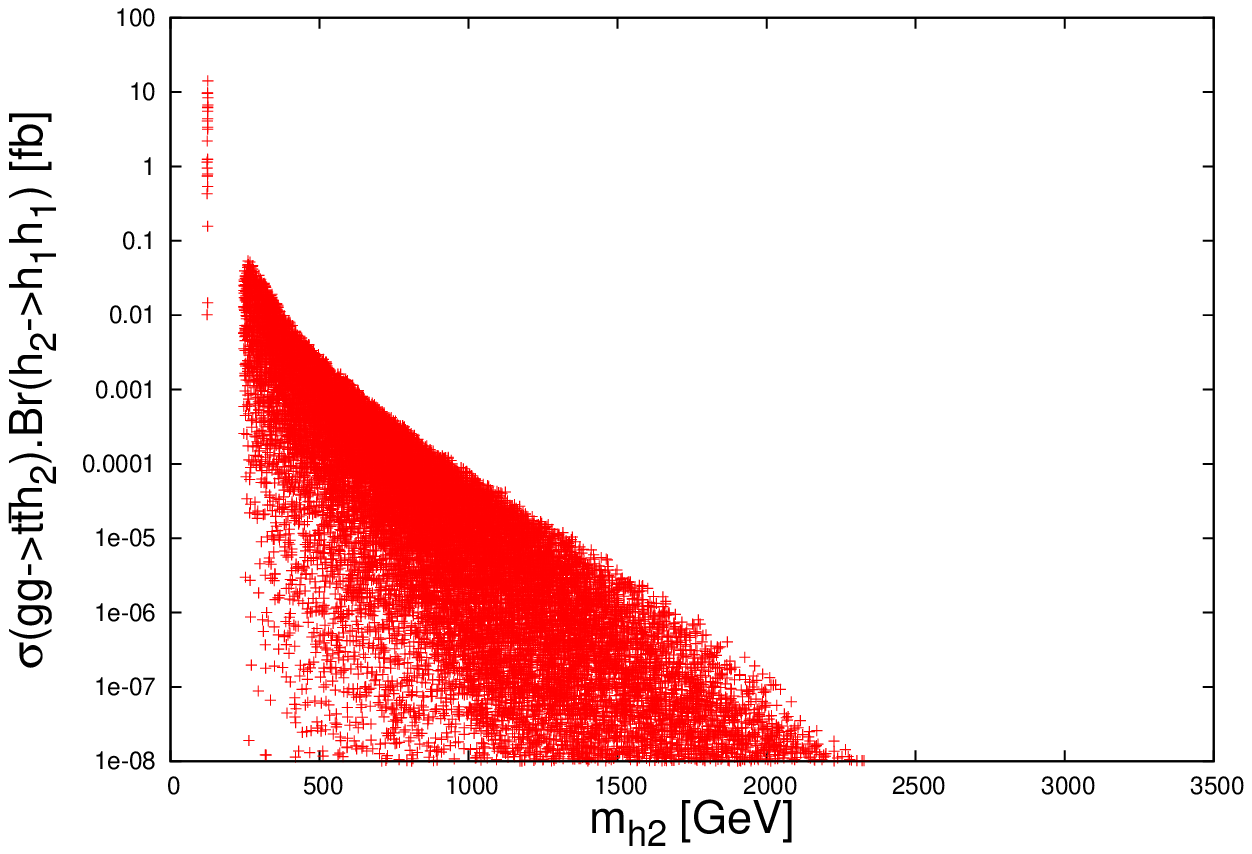}\\
    \includegraphics[scale=0.40]{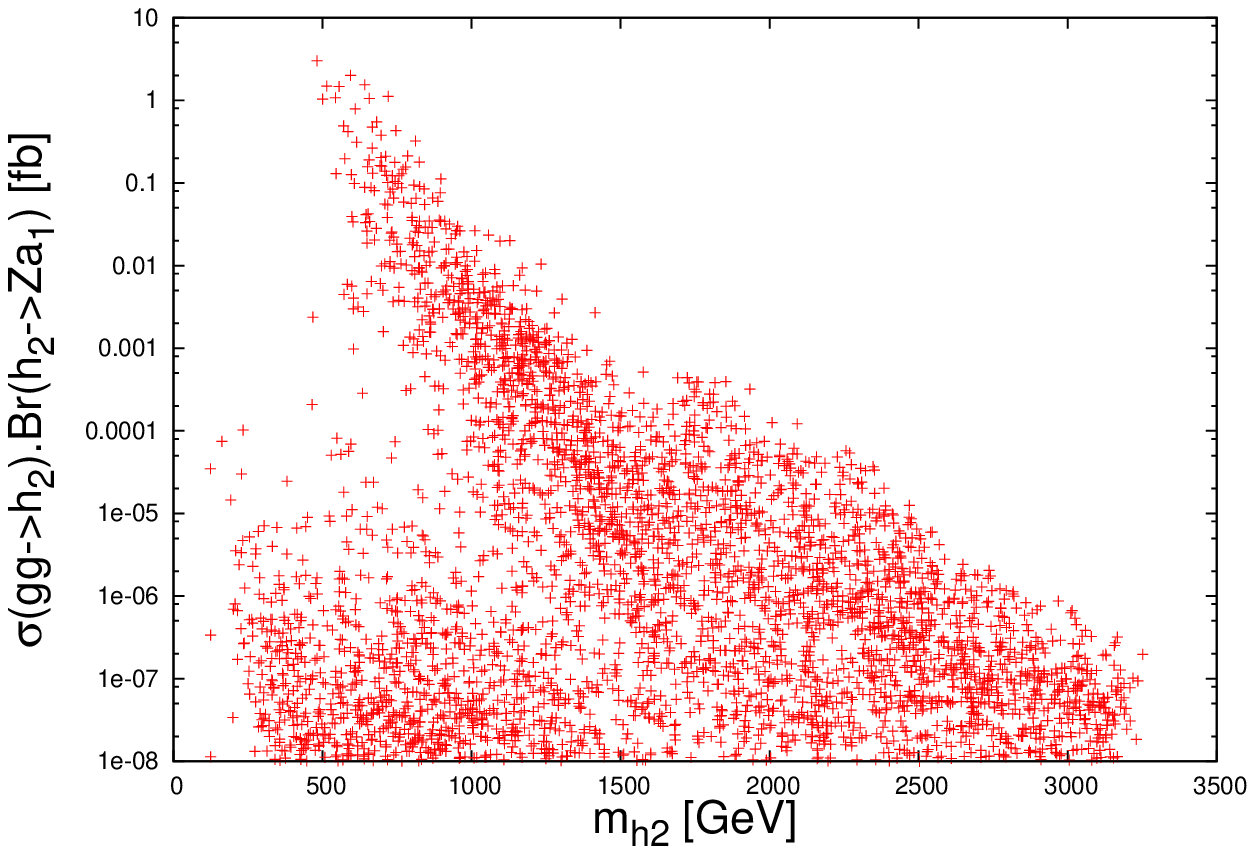}&\includegraphics[scale=0.40]{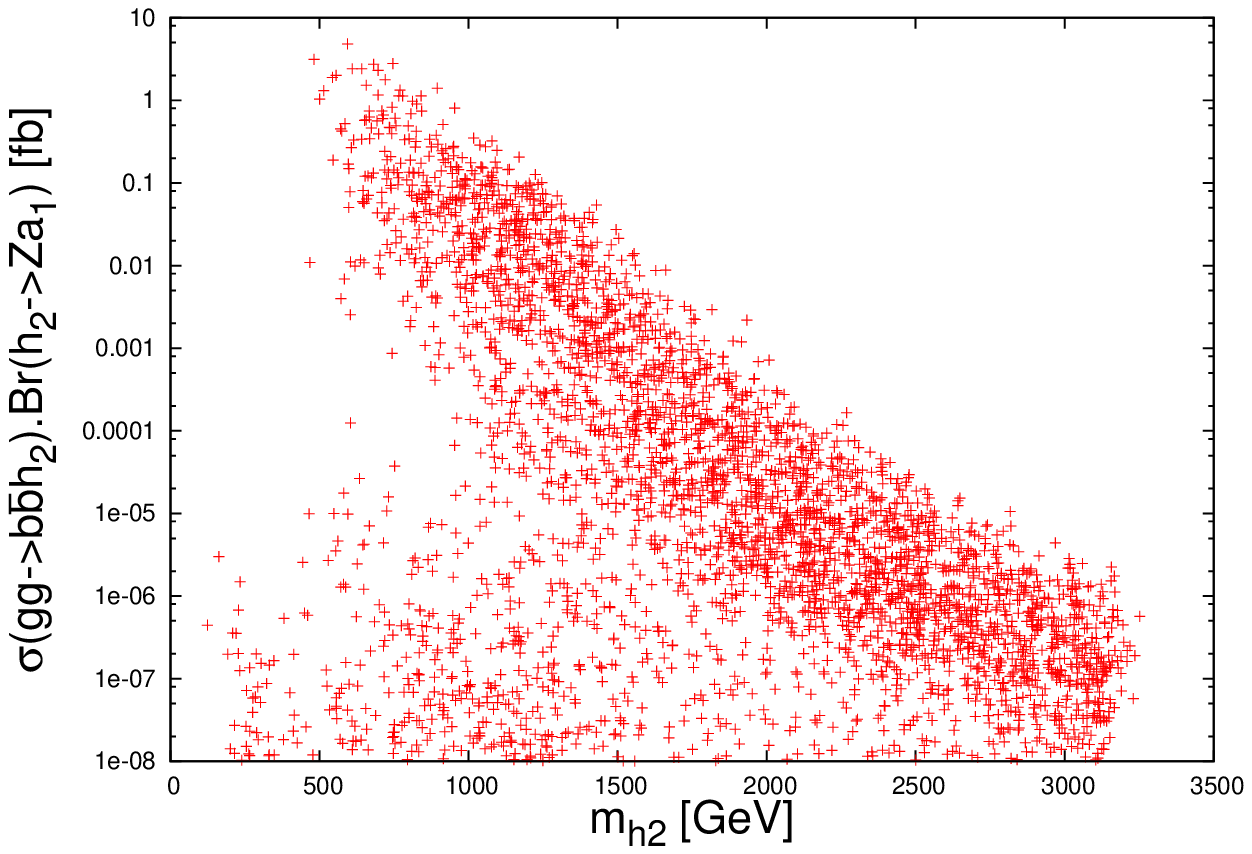}&\includegraphics[scale=0.40]{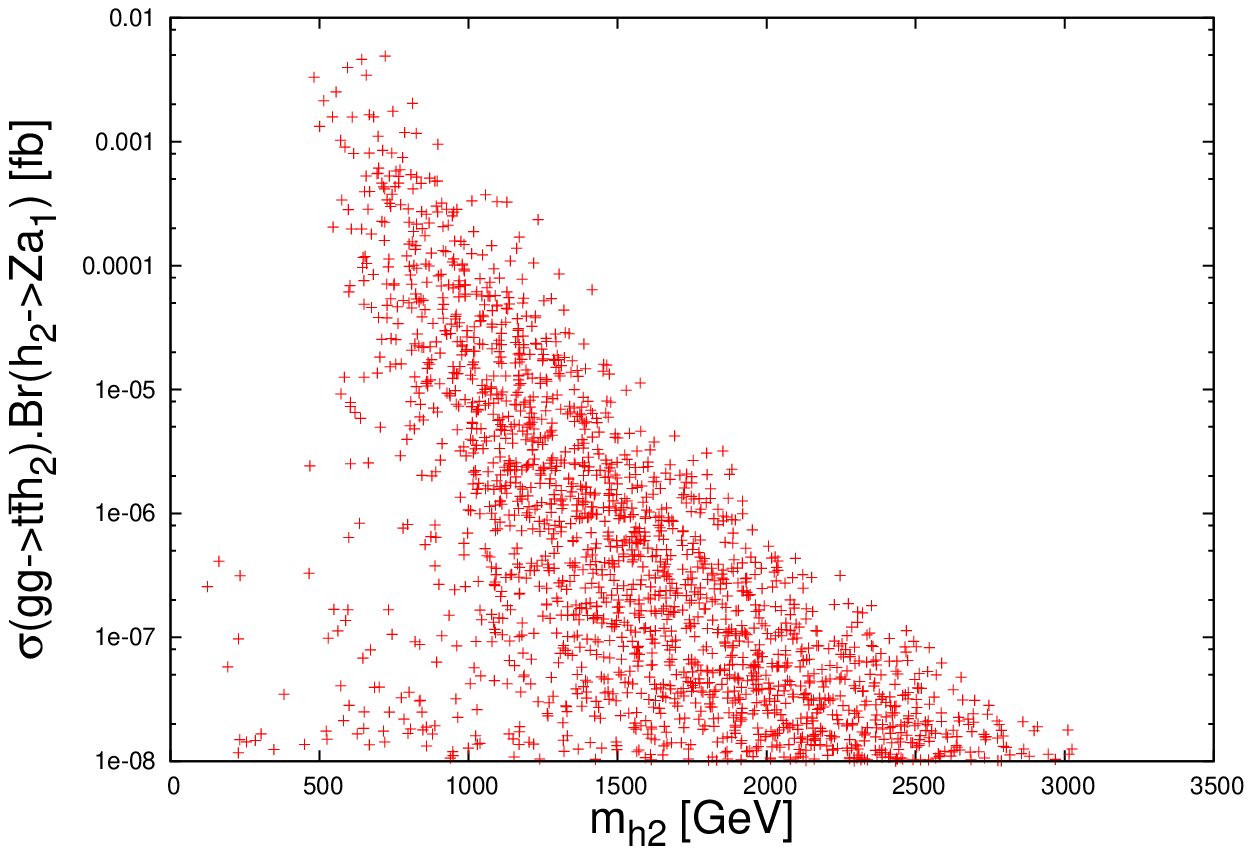}
   \end{tabular}

\caption{The signal rates for the second lightest CP-even Higgs boson $h_2$ produced in gluon fusion $\sigma(gg\to h_2)$ (left),
in association with bottom quarks $\sigma(gg\to b\bar b h_2)$ (middel) and in association with top quarks $\sigma(gg\to t\bar t h_2)$ (right)
times $~{\rm Br}(h_2\to a_1a_1)$ (top), $~{\rm Br}(h_2\to h_1h_1)$ (middle) and $~{\rm Br}(h_2\to Za_1)$ (bottom) as functions of $m_{h_2}$.}

 \label{fig5}
\end{figure}

It is interesting to note that $h_2$ decays into lighter Higgs pairs or into Higgs plus gauge boson
 are kinematically open in large areas of the NMSSM
parameter space, see Fig.~\ref{fig5} which shows the signal rates in presence of the following decays:
$h_2\to a_1a_1$ (top), $h_2\to h_1h_1$ (middle) and $h_2\to Za_1$ (bottom) for the
gluon fusion production (left), the production in association with bottom quarks (middle) and the production
in association with top quarks (right).
 It is shown from the figure that the largest
signal rates also come from the gluon fusion production channel, reaching maximum about 1000 fb 
for light $m_{h_2}$ in case of both $a_1a_1$ and $h_1h_1$ final states
and 5 fb for $m_{h_2} \sim 480$ GeV in case of $Za_1$ final state. The signal rates for
the $h_2$ production in association with both bottom and top quarks    
are quite small in most NMSSM parameter space except for small region where both the $a_1a_1$ and $h_1h_1$ final states 
reach maximum signal rates of about 20 fb for the production channel $gg\to b\bar bh_2$ (two top middle panels) and 10 fb 
for the production channel
$gg\to t\bar th_2$ (two top left panels). The $Za_1$ final state can only give sizable signal rates
in the mass range $m_{h_2} \gtrsim 480$ GeV, reaching up to 3 fb for the production channel $gg\to h_2$ (bottom left panel)
and 5 fb for the production channel $gg\to b\bar b h_2$ (bottom middle panel).
It is remarkable to notice that the $a_1a_1$, $h_1h_1$ and $Za_1$ final states will further decay into quarks or leptons. 
The promising channels which may be exploited to discover the $h_2$ 
are $2\tau 2b$ and $4\tau$ final state as the $4b$ final state may be difficult to exploit because of the large background.
Such extracted signals may require very large integrated luminosities of 300 $fb^{-1}$ or higher \footnote{The $h_2$ decays to SUSY particles, if
kinematically possible, can be dominant, making searching for $h_2$ at the LHC very complicated.}. 

\section{Conclusions}
\label{sect:summa}
The discovered SM-like Higgs boson can be accommodated in the framework of the NMSSM. In this
model, by assuming CP-conservation, there are seven Higgses: three CP-even Higgses $h_{1, 2, 3}$ ($m_{h_1} < m_{h_2} < m_{h_3}$), 
two CP-odd Higgses $a_{1, 2}$ ($m_{a_1} < m_{a_2} $) and a pair of charged Higgses $h^{\pm}$. We have computed the cross section 
for the $h_2$ produced in gluon fusion,
in association with a $b\bar b$ pair and in association with a $t\bar t$ pair in the following final states: 
 $\tau^+\tau^-$, $b\bar b$,
$t\bar t$, $\gamma\gamma$, $Z\gamma$, $W^+W^-$, $ZZ$, $a_1a_1$, $h_1h_1$ and $Za_1$ at the LHC with $\sqrt{s}$ = 14 TeV .
It has been found that the size of 
the signal rates in some regions
of the parameter space is quite large and that could help discovering the $h_2$ 
through: the $\tau^+\tau^-$, $b\bar b$, $\gamma\gamma$ and $Z\gamma$ final states in the low mass range $m_{h_2} \lesssim 2 m_{W}$.
Also, it has been found that the $W^+W^-$ and $ZZ$ final states can give considerable signal rates for the mass range $m_{h_2} \lesssim  500$ GeV.
For most of the NMSSM parameter space, the dominant contribution to the $h_2$ production comes from the gluon fusion production channel.
Furthermore, it has been noticed that 
there exist some regions of the NMSSM parameter space where  the production rates of the $h_2$ in the mass range $m_{h_2} \gtrsim 480$ GeV 
are quite sizable,
and may be exploited to discover the $h_2$ in the $\tau^+\tau^-$, $b\bar b$ and $t\bar t$ final states through its production 
in association with $b\bar b$ pair, which has the dominant contribution to the $h_2$ production in this mass range, or
through the gluon fusion channel.

Finally, we have calculated the signal rates of the $h_2$, decaying into $a_1a_1$, $h_1h_1$ and $Za_1$ final states.
Although the $h_2$ production rates are quite small in most of the NMSSM parameter space,
it has been found that there exist some regions where the signal rates of the $h_2$ are sizable and that could help discovering
the $h_2$ at the LHC in $a_1a_1$, $h_1h_1$ and $Za_1$ final states but such discovery may require very large 
integrated luminosities of 300 $fb^{-1}$ or higher.

\section*{Acknowledgments}
This work is funded by Taibah University, KSA.

\end{document}